\documentclass[fleqn,usenatbib]{mnras}

\usepackage{newtxtext,newtxmath}

\usepackage[T1]{fontenc}

\DeclareRobustCommand{\VAN}[3]{#2}
\let\VANthebibliography\thebibliography
\def\thebibliography{\DeclareRobustCommand{\VAN}[3]{##3}\VANthebibliography}

\usepackage{graphicx}    
\usepackage{amsmath}    
\usepackage{bbding}    
\usepackage{color}
\usepackage{gensymb}
\usepackage{comment}
\usepackage{empheq}
\usepackage{comment}
\usepackage{multirow}
\usepackage{threeparttable}
\usepackage{seqsplit}
\usepackage{tablefootnote}

\usepackage{scalerel,tikz}
\usetikzlibrary{svg.path}
\definecolor{orcidlogocol}{HTML}{A6CE39}
\tikzset{orcidlogo/.pic={
 \fill[orcidlogocol] svg{M256,128c0,70.7-57.3,128-128,128C57.3,256,0,198.7,0,128C0,57.3,57.3,0,128,0C198.7,0,256,57.3,256,128z};
 \fill[white] svg{M86.3,186.2H70.9V79.1h15.4v48.4V186.2z}
 svg{M108.9,79.1h41.6c39.6,0,57,28.3,57,53.6c0,27.5-21.5,53.6-56.8,53.6h-41.8V79.1z M124.3,172.4h24.5c34.9,0,42.9-26.5,42.9-39.7c0-21.5-13.7-39.7-43.7-39.7h-23.7V172.4z}
 svg{M88.7,56.8c0,5.5-4.5,10.1-10.1,10.1c-5.6,0-10.1-4.6-10.1-10.1c0-5.6,4.5-10.1,10.1-10.1C84.2,46.7,88.7,51.3,88.7,56.8z};
}}
\newcommand\orcidicon[1]{\href{https://orcid.org/#1}{\mbox{\scalerel*{
\begin{tikzpicture}[yscale=-1,transform shape]
\pic{orcidlogo};
\end{tikzpicture}
}{|}}}}

\newcommand{\aref}[1]{\hyperref[#1]{Appendix~\ref{#1}}}

\defcitealias{2018MNRAS.477.2716K}{KBFC18}

\usepackage[normalem]{ulem}

\definecolor{darkgreen}{rgb}{0.13, 0.55, 0.13}
\definecolor{brown}{rgb}{0.65, 0.16, 0.16}

\title[COLIBRE gas-phase mass-metallicity relation]{The evolution of the galaxy gas-phase mass-metallicity relation from $z=15$ to $z=0$ in the COLIBRE cosmological simulations}

\author[Sharda et al.]{Piyush Sharda$^{\orcidicon{0000-0003-3347-7094}\,1}$\thanks{sharda@strw.leidenuniv.nl (PS)}, Joop Schaye$^{\orcidicon{0000-0002-0668-5560}1}$\thanks{schaye@strw.leidenuniv.nl (JS)}, Robert J. McGibbon$^{\orcidicon{0000-0003-0651-0776}1}$, Alejandro Benítez-Llambay$^{\orcidicon{0000-0001-8261-2796}2}$,
\newauthor
Evgenii Chaikin$^{\orcidicon{0000-0003-2047-3684}3,1}$, Carlos S. Frenk$^{\orcidicon{0000-0002-2338-716X}3}$, Jacqueline Hodge$^{\orcidicon{0000-0001-6586-8845}1}$, Filip Huško$^{\orcidicon{0000-0002-1510-1731}1}$, Sylvia Ploeckinger$^{\orcidicon{0000-0002-1965-1650}4}$,
\newauthor
Alexander J. Richings$^{\orcidicon{0000-0003-0502-9235}5,6}$, and Matthieu Schaller$^{\orcidicon{0000-0002-2395-4902}7,1}$
\\
$^{1}$Leiden Observatory, Leiden University, P.O. Box 9513, 2300 RA Leiden, The Netherlands\\
$^{2}$Dipartimento di Fisica G. Occhialini, Universit`a degli Studi di Milano Bicocca, Piazza della Scienza, 3 I-20126 Milano MI, Italy\\
$^{3}$Institute for Computational Cosmology, Department of Physics, University of Durham, South Road, Durham, DH1 3LE, UK\\
$^{4}$Department of Astrophysics, University of Vienna, Türkenschanzstrasse 17, A-1180 Vienna, Austria\\
$^{5}$Centre for Data Science, Artificial Intelligence and Modelling, University of Hull, Cottingham Road, Hull, HU6 7RX, UK\\
$^{6}$E. A. Milne Centre for Astrophysics, University of Hull, Cottingham Road, Hull, HU6 7RX, UK\\
$^{7}$Lorentz Institute for Theoretical Physics, Leiden University, PO Box 9506, 2300 RA Leiden, The Netherlands\\
}
\date{Accepted XXX. Received YYY; in original form \today}

\pubyear{2026}

\begin{document}
\label{firstpage}
\pagerange{\pageref{firstpage}--\pageref{lastpage}}
\maketitle

\begin{abstract}
We present the evolution of the galaxy gas-phase mass–metallicity relation (MZR) from $z=15$ to $z=0$ in the COLIBRE cosmological hydrodynamical simulations. Amongst other novel features, COLIBRE follows the multiphase interstellar medium with gas allowed to cool to $\sim 10\,\rm{K}$, and includes a new chemistry model in which hydrogen and helium are tracked in non-equilibrium, metals are allowed to mix and diffuse, and the chemical network is coupled to a self-consistent live dust model. Using fiducial COLIBRE runs spanning particle masses from $10^5\,\rm{M_{\odot}}$ to $10^7\,\rm{M_{\odot}}$ and box sizes $25 - 400\,\rm{cMpc}$, we derive the median, mass-weighted MZRs for star-forming galaxies and compare them with a comprehensive compilation of observational data and other simulations. COLIBRE reproduces the observed MZR across cosmic time, notwithstanding the systematic uncertainties in observational measurements of the gas-phase oxygen abundances. The simulations show excellent numerical convergence and uniquely probe the full stellar mass range sampled by current observations across all redshifts. We find that the MZR is already in place at cosmic dawn ($z \approx 10$), and shows no evolution until $z \approx 5$. The slope of the MZR becomes shallower at low redshifts. The turnover at the high-mass end is largely governed by feedback from active galactic nuclei (AGN), whereas the low-mass end of the MZR sensitively depends on the strength of feedback from core collapse supernovae. Variations in the star formation efficiency or depletion of oxygen on dust grains have a more minor impact on the MZR. We identify key physical processes that shape the MZR across cosmic time and highlight where future observations can further constrain galaxy formation models.
\end{abstract}




\section{Introduction}
\label{s:intro}

Even though metals constitute at most a few per cent of the baryonic matter in the Universe, they enable us to probe cosmic structures in remarkable detail. Produced exclusively in stars and their remnants, metals are nevertheless found ubiquitously across a wide range of environments, from planetary cores to the intra-cluster medium. Because their production is tightly linked to stellar evolution, the distribution and abundance of metals encode valuable information about the build-up of stars and galaxies over cosmic time. This makes metal abundances a powerful tracer of galaxy formation and evolution, effectively providing a fossil record of how galaxies assemble their baryonic mass. Oxygen, in particular, owing to its high cosmic abundance and the relative ease with which its emission lines can be observed, has become the primary tracer of gas-phase metallicity in galaxies \citep[see reviews by][]{2019ARA&A..57..511K,2019A&ARv..27....3M}. Thanks to the wealth of observational studies over the past three decades, it is now well established that more massive galaxies are more metal-enriched than their lower-mass counterparts, giving rise to the galaxy mass–metallicity relation \citep[MZR,][]{1989ApJ...347..875S,1994ApJ...420...87Z,2004ApJ...613..898T,2010MNRAS.408.2115M}.\footnote{In this work, MZR exclusively refers to the gas-phase mass-metallicity relation.}

The MZR has emerged as a cornerstone of galaxy evolution studies and is now recognised as one of the fundamental galaxy scaling relations across cosmic time. As a result, both observational efforts \citep[e.g.,][]{2004ApJ...613..898T,2013A&A...549A..25P,2017MNRAS.469.2121S,2019MNRAS.484.3042S,2020MNRAS.491..944C} and theoretical or numerical models \citep[e.g.,][]{2008MNRAS.387..577O,2016MNRAS.456.2140M,2017MNRAS.472.3354D,2019MNRAS.484.5587T,2025A&A...699A...6P,2026MNRAS.tmp...20M} routinely use the MZR as a key diagnostic to test our understanding of galaxy formation. By linking stellar mass to metal enrichment, the MZR provides a powerful framework for reconstructing the baryon cycle in galaxies. In particular, it encodes the cumulative effects of gas accretion, star formation, metal production, and feedback-driven outflows, thereby offering critical insight into the physical processes that regulate galaxy growth \citep[e.g.,][]{2007ApJ...658..941D,2008MNRAS.385.2181F,2011MNRAS.417.2962P,2012MNRAS.421...98D,2013ApJ...772..119L,2019MNRAS.487..456B,2021MNRAS.502.5935S,2024MNRAS.528.2232S}.

The origin of the MZR and the correlation of its slope and scatter with galaxy properties remain a topic of active investigation. For example, in addition to stellar mass, galaxy metallicity has also been found to correlate with the star formation rate \citep[SFR,][]{2008ApJ...672L.107E,2010MNRAS.408.2115M,2010A&A...521L..53L,2013ApJ...765..140A}; gas fraction \citep[][]{2013MNRAS.433.1425B,2014A&A...563A..58T,2015ApJ...812...98J,2015MNRAS.452..486D,2016MNRAS.459.2632L}; specific star formation rate \citep[sSFR, defined as the ratio of the SFR to stellar mass,][]{2013MNRAS.433L..35L,2014ApJ...797..126S,2019A&A...627A..42C}; galactic outflows \citep[][]{2011MNRAS.417.2962P,2018MNRAS.481.1690C,2021MNRAS.504...53S}; active galactic nuclei \citep[AGN,][]{2018A&A...616L...4M,2019MNRAS.486.5853D,2023MNRAS.520.1687A,2024MNRAS.529.4993L}; and the large-scale galaxy environment \citep[e.g.,][]{2012ApJ...749..133P,2017MNRAS.468.1881W,2026arXiv260303951R,2026arXiv260425632L,2026arXiv260525557B}. While some studies argue that stellar mass is the fundamental quantity that drives the MZR \citep[e.g.,][]{2023MNRAS.521.4173B,2023MNRAS.519.1149B}, others suggest it is the galactic potential, gas fraction and/or SFR \citep[e.g.,][]{2016MNRAS.459.2632L,2016MNRAS.456.1235S,2024A&A...681A.121S,2024A&A...682L..11S}, with some also finding it depends on where in the galaxy the metallicity is measured \citep[e.g.,][]{2021MNRAS.504...53S,2026MNRAS.545f2011K}. 

However, gas-phase metallicity measurements are subject to a range of systematic uncertainties. Some of these are physical in origin, such as limitations in photoionization modeling, which often assume constant temperature/density in H\textsc{ii} regions \citep[e.g.,][]{2001ApJ...556..121K,2006MNRAS.372..961K,2022ApJ...934L...8J,2023MNRAS.522L..89C}, uncertainties in calibration of emission line ratios due to degeneracies involving gas temperature, density and pressure \citep[e.g.,][]{2008ApJ...681.1183K,2009MNRAS.398..485P,2019MNRAS.487...79P,2025ApJ...995..204M}, ionization state \citep[e.g.,][]{2002ApJS..142...35K,2012ApJ...757..108Y,2019MNRAS.487..333H}, turbulence and shocks \citep[e.g.,][]{2016MNRAS.455.3058M,2019MNRAS.485L..38D,2025ApJ...988..261Z}, and sensitivity to abundances of other elements like N \citep[e.g.,][]{2009MNRAS.398..949P,2017ApJ...836..164S,2025arXiv250810099S,2026ApJ...997L..44R}. Other systematics arise from observational and methodological issues, such as sample inhomogeneity, aperture and S/N effects, corrections for unseen ionization states, limited availability of more robust metallicity diagnostics, and contamination from diffuse ionized gas \citep[e.g.,][]{2008MNRAS.383..209H,2010ApJ...720.1738P,2014ApJ...797..126S,2017ApJ...850..136S,2017MNRAS.466.3217Z,2018MNRAS.474.3727L,2019MNRAS.485..367K,2020MNRAS.495.3819A}. As a result, the normalization of the MZR remains poorly constrained, with systematic uncertainties of up to $\pm0.2\,\rm{dex}$ in metallicity measurements across large galaxy samples being difficult to avoid. While methods that can disentangle the ionization state of the gas from its metallicity -- such as recombination and auroral line diagnostics -- provide more robust abundance measurements, the intrinsic faintness of the required emission lines limits the size of samples to which these techniques can be applied \citep[e.g.,][]{2017PASP..129h2001P,2017MNRAS.466.4403N}.

Nevertheless, the last decade has witnessed a surge in measurements of the MZR, even out to high redshifts, with several large-scale ground-based surveys \citep[e.g.,][]{2006ApJ...644..813E,2014ApJ...792....3M,2014A&A...563A..58T,2014ApJ...792...75Z,2015ApJ...801...88S,2015ApJ...799..209W,2021MNRAS.500.4229G,2025MNRAS.544.4025L,2025ApJ...984..188K}. Recent observations from the \textit{James Webb Space Telescope} (JWST) have pushed the frontier to enable metallicity measurements of the first galaxies within a few $100\,\rm{Myr}$ of the Big Bang, revealing the presence of metal-rich galaxies in the first Gyr \citep[e.g.,][]{2023ApJS..269...33N,2024AA...684A..75C,2026AA...706A.165K,2026MNRAS.546f2023R,2026OJAp....956033N}. On the theoretical front, both models and simulations have offered a variety of explanations for the existence of the MZR at the highest redshifts \citep[e.g.,][]{2020MNRAS.494.1988L,2021A&A...651A.109D,2023MNRAS.518.3557U,2023MNRAS.519.3118W,2024MNRAS.527.5929Y,2024ApJ...967L..41M,2025MNRAS.536..119G,2025A&A...699A...6P}. Despite the systematic uncertainties discussed above, both observations and theory find an evolution in the slope and normalization of the MZR with redshift \citep[][see their figure 25]{2019A&ARv..27....3M}. These observations raise key questions on early galaxy evolution and chemical enrichment: how did galaxies become so metal-rich so early on? Was the MZR already in place at cosmic dawn ($z \approx 10$), and has it evolved across cosmic time? How were metals transported outside the galaxies to enrich the intergalactic medium? Does galaxy metallicity correlate with other galaxy properties in similar ways across cosmic time?

In this work, we study the evolution of the MZR across cosmic time in the COLIBRE\footnote{The acronym COLIBRE stands for \textbf{COL}d \textbf{I}sm and \textbf{B}etter \textbf{RE}solution.} cosmological smoothed particle hydrodynamical (SPH) simulations \citep[][]{2025arXiv250821126S,chaikin25a}. The key improvements that make COLIBRE simulations stand out in comparison to previous simulations, particularly for studies of galactic chemical evolution, include:

\begin{itemize}
\item Explicitly following the cold dense phase of the interstellar medium (ISM) by removing the pressure floor and allowing the gas to cool down to $\approx 10\,\rm{K}$. 
\item Tracing non-equilibrium chemistry of 10 hydrogen and helium species (including self-shielding and local radiation sources) and tracking abundances of over 10 different elements, including $r-$ and $s$- process elements.
\item Incorporating a \textit{live} dust model with varying dust compositions and grain size distributions that is self-consistently coupled to the thermochemistry of the gas \citep[][]{2025MNRAS.543..891P,2026MNRAS.545f2040T}. 
\item Including a sub-grid model for stellar mass loss using up to date nucleosynthetic yields and unresolved small-scale turbulent metal mixing and diffusion \citep[][]{2026arXiv260400980C} that can impact both integrated and spatially-resolved gas-phase metallicities \citep[e.g.,][]{2012ApJ...758...48Y,2018MNRAS.474.2194E,2021MNRAS.502.5935S}.
\item Including a model for pre-supernova feedback \citep[][]{2026MNRAS.546ag268B} that can have important implications for the chemical enrichment of the ISM \citep[e.g.,][]{2018MNRAS.479.1866K,2022MNRAS.509..272C}. 
\end{itemize}

The success of COLIBRE in reproducing a variety of galaxy scaling relations across cosmic times has been documented in several works, such as the galaxy stellar, gas (H\textsc{i} and H$_2$) and dust mass functions, present-day gas-phase and stellar MZR \citep[][]{2025arXiv250821126S}, the star-forming main sequence and the stellar mass function across cosmic time \citep[][]{chaikin25b}, the evolution of galaxy size and angular momentum scaling relations \citep[][]{2026arXiv260326200L}, star formation and Kennicutt-Schmidt relations \citep[][]{2025arXiv251211309L}, number density of massive quiescent galaxies at high redshifts \citep[][]{2025arXiv251216208C}, and the far-UV to near IR and far IR to sub-mm UV luminosity function at $z=0$ \citep[][]{2026arXiv260502022L}. We show in this work that COLIBRE reproduces the MZR across cosmic time, showing good agreement with observations at the highest redshifts, where theoretical efforts have traditionally fallen short of matching the observed metallicities and their rapid early build-up.

We organize this paper as follows: \autoref{s:setup} provides a brief overview of the COLIBRE simulations suite, particularly focusing on the gas-phase oxygen abundance measurements, \autoref{s:results} presents the predicted MZRs, and \autoref{s:discussion} discusses the impact of measurement choices and variations in the COLIBRE galaxy formation model on the MZR. Finally, we summarize our findings in \autoref{s:summary}. The COLIBRE Project uses the following $\Lambda$CDM parameters: $\Omega_{\rm{m,0}}$ = 0.306, $\Omega_{\rm{b,0}}$ = 0.0486, $\sigma_8$ = 0.807, $h$ = 0.681, and $n_{s}$ = 0.967 \citep[][]{2022PhRvD.105b3520A}. The Solar abundance of oxygen is set to $12 + \log_{10}\rm{O/H} = 8.69$ \citep[][]{2009ARA&A..47..481A}.

\section{COLIBRE Simulations}
\label{s:setup}

The initial conditions for the COLIBRE simulations were generated at $z=63$ \citep[see section 2.2 of][]{2025arXiv250821126S} using \texttt{MONOFONIC} \citep[][]{2021MNRAS.503..426H,2021MNRAS.500..663M}. The initial mass fractions are $X=0.756$ for hydrogen and $Y=0.244$ for helium. The simulations were run using the \texttt{SWIFT} code \citep{2024MNRAS.530.2378S}, employing the \texttt{SPHENIX} SPH scheme for hydrodynamics \citep[][]{2022MNRAS.511.2367B}. Radiative cooling and chemistry are followed using a hybrid version of the \texttt{CHIMES} network \citep[][]{2014MNRAS.440.3349R,2014MNRAS.442.2780R,2025MNRAS.543..891P}, with hydrogen and helium species followed without assuming equilibrium, while metals use tabulated equilibrium rates corrected for the non-equilibrium abundance of free electrons from hydrogen and helium. Star formation follows a thermal-turbulent velocity-based gravitational instability criterion \citep{2024MNRAS.532.3299N}, with gas particles satisfying this criterion converted into stars according to the volumetric \citet{1959ApJ...129..243S} law. Supernova feedback follows the thermal-kinetic subgrid model introduced in \citet{2023MNRAS.523.3709C}, with adjustments as described in \citet[][section 3.7]{2025arXiv250821126S}. Feedback from AGN is implemented as thermal energy injection into gas particles neighbouring the supermassive black hole (\citealt{2009MNRAS.398...53B}; see also \autoref{s:agn}, where we assess the MZR from a `hybrid' AGN feedback model introduced in \citealt{2026MNRAS.547ag324H}). COLIBRE adopts the default metal diffusion model presented in \citet{2026arXiv260400980C} , noting that reduced diffusion efficiencies lead to steeper MZRs by increasing the metallicity at the high-mass end by up to $0.3\,\rm{dex}$ at $z=0$ \citep[][see their figure~6]{2026arXiv260400980C}.

Chemical enrichment of the interstellar medium (ISM) follows the model presented in \citet{2026arXiv260400980C}, which is based on \citet{2009MNRAS.399..574W}, and includes contributions from asymptotic giant branch (AGB) stars, massive stars, and both core-collapse and Type Ia supernovae. Each star particle represents a single stellar population, with zero-age main sequence masses drawn from the \citet{2003PASP..115..763C} initial mass function (IMF) over the range $0.1$--$100\,\rm{M_{\odot}}$. Stellar metallicities are inherited from the parent gas particle, including both diffused and solid-phase abundances. Nucleosynthetic yields for AGB stars are taken from \citet{2010MNRAS.403.1413K,2014MNRAS.437..195D,2014ApJ...797...44F,2016ApJ...825...26K} and \citet{2022MNRAS.510.1557C}, for pre-supernova mass loss from \citet{2006ApJ...653.1145K}, and for supernovae from \citet{2013ARA&A..51..457N}.

The COLIBRE simulations are performed in a `wedding cake' configuration, with multiple box sizes at three resolutions (m5, m6, and m7), corresponding to mean initial particle masses of $\approx 10^5$, $10^6$, and $10^7\,\rm{M_{\odot}}$ for both baryons and dark matter, respectively. At each resolution, the simulations employ four times as many dark matter particles as baryons to minimise spurious energy transfer from dark matter to stellar particles \citep[][]{2023MNRAS.525.5614L}. Galaxies are identified using the \texttt{HBT-HERONS} subhalo finder \citep[][]{2018MNRAS.474..604H,2025MNRAS.543.1339F}. \autoref{tab:tab1} summarises the key characteristics of the fiducial simulations we use in this work; we also utilise smaller boxes in the COLIBRE suite when discussing the impact of physical and modeling choices on the MZR later in \autoref{s:discussion}. As the largest boxes at m5 resolution have not yet reached $z=0$, we use the largest available box at each redshift. The simulations are calibrated to reproduce the $z=0$ galaxy stellar mass function \citep[][]{2022MNRAS.513..439D}, galaxy mass -- size function \citep[][]{2022MNRAS.509.3751H}, and black hole -- stellar mass function \citep[][]{2023MNRAS.518.2177G,2024MNRAS.530.3429G}. We refer the reader to \citet{2025arXiv250821126S} for a comprehensive overview of the COLIBRE project, and \citet{chaikin25a} for further details on the calibration strategy.

\begin{table}
\centering
\caption{Characteristics of the different (fiducial) COLIBRE simulations used in this work. From left to right, the columns show the simulation identifier, the length of the simulation box (in comoving Mpc), the number of baryonic particles, the baryon and dark matter particle masses, and the redshift range over which the simulation box is used to study the MZR.}
\begin{tabular}{|l|r|r|l|l|r|}
\hline
Identifier & $L/\rm{cMpc}$ & $N_{\rm{b}}$ & $m_{\rm{b}}/\rm{M_{\odot}}$ & $m_{\rm{DM}}/\rm{M_{\odot}}$ & $z$ \\
\hline
\hline
L025m5 & 25 & $752^3$ & $2.30\times10^5$ & $3.03\times10^5$ & $0$ \\
L050m5 & 50 & $1504^3$ & $2.30\times10^5$ & $3.03\times10^5$ & $1 - 2$ \\
L100m5 & 100 & $3008^3$ & $2.30\times10^5$ & $3.03\times10^5$ & $\geq 3$ \\
\hline
L200m6 & 200 & $3008^3$ & $1.84\times10^6$ & $2.42\times10^6$ & $0 - 15$ \\
\hline
L400m7 & 400 & $3008^3$ & $1.47\times10^7$ & $1.94\times10^7$ & $0 - 15$ \\
\hline
\label{tab:tab1}
\end{tabular}
\end{table}

\subsection{Metallicity estimation}
\label{s:metallicity_estimation}
To estimate the galaxy metallicity, unless otherwise specified, we only consider gas that is cool ($T < 10^{4.5}\,\rm{K}$) and dense ($n_{\rm{H}} > 0.1\,\rm{cm^{-3}}$, where $n_{\rm{H}}$ is the ratio of gas volume density to the mass of hydrogen atom). We explore the impact of using alternate selection criteria to estimate the metallicity in \autoref{s:density}. We use the mass-weighted ratio of the abundances of oxygen over hydrogen in the gas. Unlike previous generations of cosmological simulations, we account for oxygen depleted onto dust grains. For each gas particle $i$ associated with a galaxy, we calculate the mass-weighted metallicity of the galaxy as
\begin{equation}
12 + \log_{10} {\rm O/H} = 12 + \log_{10} \left( \frac{\sum_i m_i (n_{{\rm O},i} / n_{{\rm H},i})}{\sum_i m_i} \right)
\label{eq:oxygen}
\end{equation}
where $m_i$ is the mass of the $i^{\rm{th}}$ particle, and $n_{\mathrm{O},i}$ and $n_{\mathrm{H},i}$ are the number densities of gas-phase oxygen and hydrogen nuclei for the $i^{\rm{th}}$ particle, respectively.

\begin{figure*}
\centering
\includegraphics[width=\textwidth]{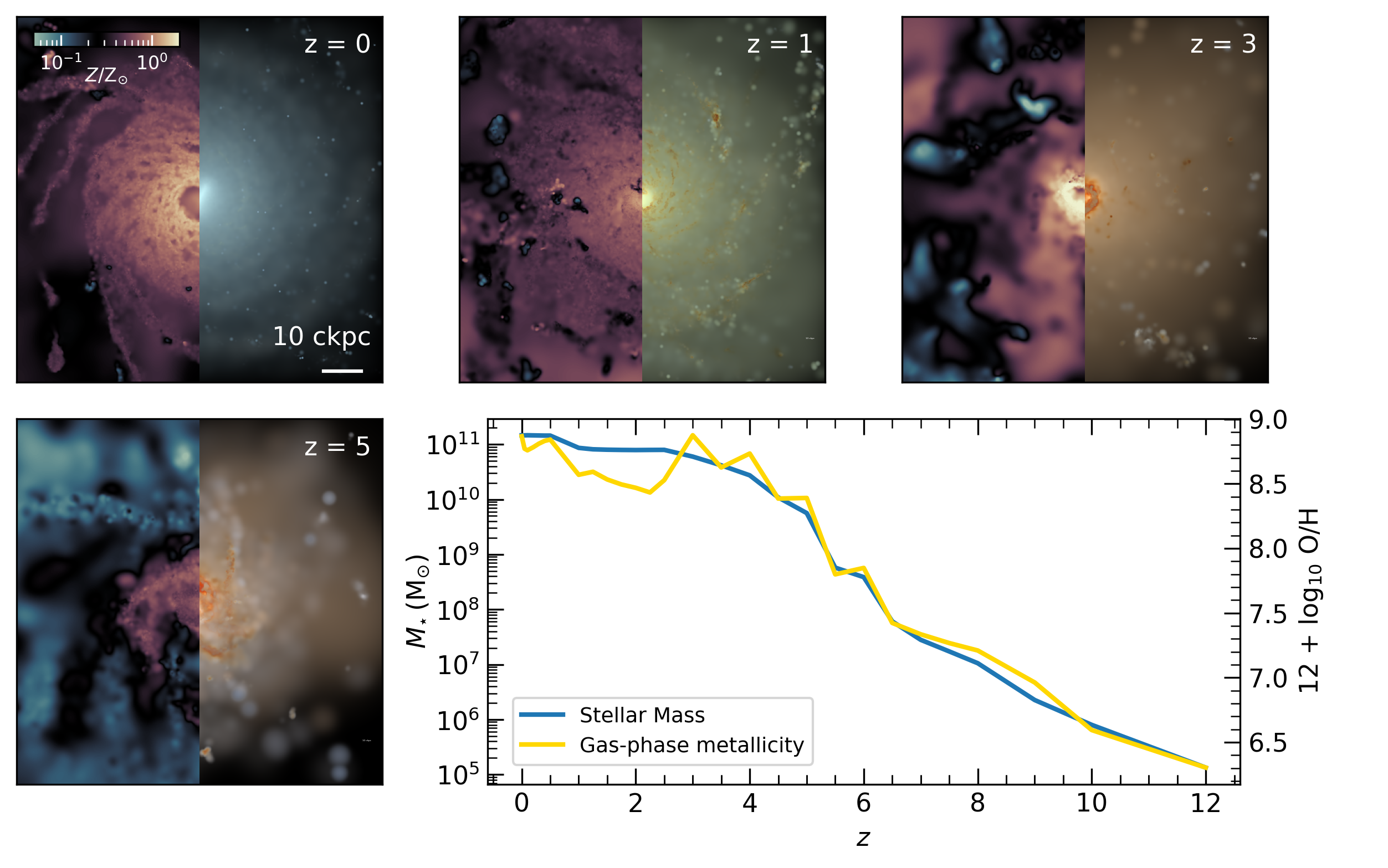}
\caption{Face-on projections within a 50 comoving kpc volume of the gas-phase metallicity (left half of the image in each redshift panel) and integrated stellar light (right half of the image) as would be seen through near-IR JWST filters across cosmic time in a representative massive spiral galaxy from the COLIBRE L025m5 simulation. The metallicity scale is depicted in the top left panel, and stretches from $0.05\,\rm{Z_{\odot}}$ to $2\,\rm{Z_{\odot}}$. The RGB colors for the post-processed stellar light images correspond to central (observed-frame) wavelengths $4.44\,\mu\rm{m},\,3.56\,\mu\rm{m}$ and $2.57\,\mu\rm{m}$, respectively. The bottom right panel shows the evolution of stellar mass and metallicity across cosmic time. The metallicity increases non-monotonically due to accretion and outflows, as the galaxy grows and evolves to a $10^{11.16}\,\rm{M_{\odot}}$ spiral by $z = 0$.}
\label{fig:visual}
\end{figure*}

To further select only star-forming galaxies, following \citet[section 3.5]{chaikin25b}, we apply a cut on the galaxy sSFR, excluding galaxies with sSFR $< 0.2/t_{\rm{H}}(z)$, where $t_{\rm{H}}(z)$ is the Hubble time at redshift $z$. This criterion reduces to classifying $z=0$ galaxies with sSFR $\gtrsim 0.01\,\rm{Gyr^{-1}}$ as star-forming, a threshold commonly used in earlier works \citep[e.g.,][]{2013ApJ...778..126B,2021MNRAS.500.2036K}, and ensures that the sSFR threshold increases with redshift as observations of gas-phase metallicity tend to be limited to actively star-forming galaxies at high $z$ \citep[e.g.,][]{2022MNRAS.516..975T}. Selecting galaxies above a constant sSFR = $0.01\,\rm{Gyr^{-1}}$ instead yields a larger sample, particularly at high redshifts, but the median trends remain similar to our default classification above. Unless otherwise specified, we calculate the stellar mass and SFR from particles bound to the galaxy subhalo that lie within a $50\,\rm{kpc}$ aperture centred at the location of the most bound particle. For the oxygen abundance, we use a smaller aperture (physical size 3 kpc), in line with previous works \citep[e.g.,][]{2023MNRAS.522.3831F}, to mimic the SDSS fibre size at $z \approx 0$. We discuss the impact of aperture sizes on the MZR in \autoref{s:apertures}.

\begin{table*}
\centering
\caption{Summary of observational datasets included in this work. From left to right, the columns show the serial number, name or identifier of the dataset, redshift range covered, number of galaxies, emission line(s) used to derive the oxygen abundance (or SED fitting), and associated reference(s). Datasets that use at least one auroral emission line are highlighted in bold. Note that the emission lines listed for a sample are not necessarily available for all galaxies in that sample, and doublets are not always resolved into individual components. For large datasets (marked by $\star$), only the median abundances are used to create the MZR. The diamond symbol ($\Diamond$) implies only stacked measurements are used to infer metallicities.}
\begin{tabular}{|l|l|r|r|l|r|}
\hline
No. & Dataset & $z$ & $N_{\rm{gal}}$ & Emission Line(s) & Reference \\
\hline
\hline
1 & SDSS DR2$^{\star}$ & 0.005 -- 0.30 & 53400  & [O\textsc{ii}]$\lambda\lambda 3727, 3729$, H$\beta$, [O\textsc{iii}]$\lambda\lambda 4959, 5007$, & \protect\citet{2004ApJ...613..898T} \\
& & & & H$\alpha$, [N\textsc{ii}]$\lambda 6584$, [S\textsc{ii}]$\lambda\lambda 6717, 6731$ & \\
2 & SDSS DR7$^*$ & 0.03 -- 0.30 & 153452 & Same as 1 & \protect\citet{2020MNRAS.491..944C} \\
3 & \textbf{Nearby dwarfs} & $\ll 0.01$ & 25  & [O\textsc{ii}]$\lambda\lambda 3727, 3729$, \textbf{[O\textsc{iii}]$\lambda 4363$,} [N\textsc{ii}]$\lambda 6584$ & \citet{2006ApJ...647..970L} \\
4 & SAMI DR3$^{\star}$ & 0.04 -- 0.1 & 472  & [O\textsc{ii}]$\lambda\lambda 3727, 3729$, H$\beta$, [O\textsc{iii}]$\lambda\lambda 4959, 5007$, & \protect\citet{2022MNRAS.510..320F} \\
& & & & [N\textsc{ii}]$\lambda\lambda 6548, 6584$ & \\
5 & \textbf{SDSS MaNGA}$^*$ & 0.01 -- 0.25 & 264  & \textbf{[O\textsc{iii}]$\lambda 4363$,} \textbf{[O\textsc{ii}]$\lambda\lambda 7322, 7332$}, \textbf{[N\textsc{ii}]$\lambda 5755$} & \citet{2020AA...634A.107Y} \\
6 & \textbf{MUDF} & 0.3 -- 2.4 & 70 & \textbf{[O\textsc{iii}]$\lambda\lambda 1661, 1666$}, [O\textsc{ii}]$\lambda\lambda 3727, 3729$, H$\beta$, & \citet{2024ApJ...966..228R} \\
& & & & \textbf{[O\textsc{iii}]$\lambda 4363$}, [O\textsc{iii}]$\lambda\lambda 4959, 5007$ & \\
7 & LEGA-C & 0.6 -- 0.8 & 145 & [O\textsc{ii}]$\lambda\lambda 3727, 3729$, H$\beta$, [O\textsc{iii}]$\lambda\lambda 4959, 5007$ & \citet{2024ApJ...964...59L} \\
8 & KROSS+KGES$^{\star}$ & 0.6 -- 1.8 & 644  & [N\textsc{ii}]$\lambda 6584$, H$\alpha$ & \citet{2021MNRAS.500.4229G} \\
9 & KMOS3D & 0.6 -- 2.7 & 259  & Same as 8 & \citet{2016ApJ...827...74W} \\
10 & DEEP2$^{\star}$ & 0.7 -- 1.4 & 1350 & Same as 7 & \protect\citet{2011ApJ...730..137Z} \\
11 & NGDEEP$^{\Diamond}$ & 1.1 -- 3.4 & Stack & Same as 1 + [Ne\textsc{iii}]$\lambda 3870$ & \citet{2026arXiv260520810H} \\
12 & CANDELS & 1.2 -- 1.5 & 49 & Same as 7 & \citet{2021ApJ...919..143H} \\
13 & JADES + DarkHorse + OASIS & 1.2 -- 9.1 & 50 & Same as 7 + [Ne\textsc{iii}]$\lambda 3870$ & \citet{2026arXiv260611345I} \\
14 & MOSDEF$^{\star}$ & 1.3 -- 3.3 & 295 & Same as 7 & \protect\citet{2026ApJ..1000..109J} \\
15 & SINS/zc-SINF & 1.4 -- 2.5 & 24  & Same as 8 & \citet{2018ApJS..238...21F} \\
16 & \textbf{AURORA} & 1.4 -- 7.2 & 41 &  \textbf{[S\textsc{ii}]$\lambda 4070$,} \textbf{[O\textsc{iii}]$\lambda 4363$,} \textbf{[S\textsc{iii}]$\lambda 6314$,} & \citet{2025arXiv250810099S}, \\
& & & & [O\textsc{ii}]$\lambda\lambda 7322, 7332$ & \citet{2025arXiv251216989K} \\
17 & EXCELS & 1.6 -- 7.9 & 65  & Same as 13 & \citet{2026MNRAS.tmp..397S} \\
18 & A2744+SMACS & 1.8 -- 3.3 & 51 & Same as 7 & \citet{2023ApJ...955L..18L} \\
19 & \textbf{MARTA} & 1.8 -- 4.7 & 16 & [S\textsc{ii}]$\lambda\lambda 4068, 4076$, \textbf{[O\textsc{iii}]$\lambda 4363$,} \textbf{[S\textsc{iii}]$\lambda 6312$,} & \citet{2025AA...703A.208C} \\
& & & & [O\textsc{ii}]$\lambda\lambda 7322, 7332$ & \\
20 & CECILIA & 2.1 -- 2.6 & 7  & [O\textsc{iii}]$\lambda\lambda 4959, 5007$, H$\beta$ & \citet{2025arXiv251200162R} \\
21 & KBSS$^*$ & 2.1 -- 2.6 & 243 & Same as 8 + 20 & \citet{2014ApJ...795..165S} \\
22 & NIRVANDELS & 3.0 -- 3.7 & 21  & Same as 13 & \citet{2024MNRAS.532.3102S} \\
23 & JADES & 3.0 -- 9.4 & 77  & Same as 13 & \citet{2024AA...684A..75C} \\
24 & \textbf{CEERS} & 3.8 -- 8.9 & 133  & Same as 13 + \textbf{[O\textsc{iii}]$\lambda 4363$} & \citet{2023ApJS..269...33N} \\
25 & \textbf{GLASS} & 3.9 -- 7.9 & 14 & Same as 24 & \citet{2023ApJS..269...33N} \\
26 & \textbf{ALPINE-CRISTAL} & 4.4 -- 5.7 & 18 & Same as 7 + \textbf{[O\textsc{iii}]$\lambda 4363$} + [O\textsc{ii}]$\lambda\lambda 7322, 7332$ & \citet{2026ApJS..282...19F} \\
27 & \textbf{ERO} & 5.3 -- 8.5 & 5 & Same as 24 & \citet{2023ApJS..269...33N} \\
28 & SAPPHIRES & 5.8 -- 6.8 & 7 & Same as 20 & \citet{2025arXiv250503873H} \\
29 & \textbf{EIGER+ALT+COLA1}$^{\Diamond}$ & 6.0 -- 7.0 & Stack & \textbf{[O\textsc{iii}]$\lambda 4363$}, H$\beta$, [O\textsc{iii}]$\lambda\lambda 4959, 5007$ & \citet{2026AA...706A.165K} \\
30 & PRIMAL & 6.0 -- 9.4 & 27  & Same as 1 & \protect\citet{2025ApJ...978..136S} \\
31 & \textbf{GLIMPSE-D} & 6.1 -- 7.6 & 8 & Same as 7 + \textbf{[O\textsc{iii}]$\lambda 4363$} + H$\alpha$ & \citet{2026arXiv260506770H} \\
32 & UNCOVER & 6.2 -- 7.7 & 7 & Same as 7 & \protect\citet{2024ApJ...976L..15C} \\
33 & REBELS & 6.5 -- 7.7 & 12  & Same as 11 & \protect\citet{2026MNRAS.546f2023R} \\
34 & \textbf{New + literature} & 7.7 -- 10.0 & 10  & Same as 7 + \textbf{[O\textsc{iii}]$\lambda 4363$} + [O\textsc{iii}]$\, 88\mu\rm{m}$ & \citet{2023ApJ...957...39L} \\
35 & CANUCS-A370-z8-LAE & 8.2 & 1 & Same as 20 & \citet{2025ApJ...988...26W} \\
36 & \textbf{Firefly Sparkle} & 8.3 & 1 & Same as 7 + [Ne\textsc{iii}]$\lambda 3870$ + \textbf{[O\textsc{iii}]$\lambda 4363$} + & \citet{2024Natur.636..332M} \\
 &  &  &  & \textbf{[O\textsc{iii}]$\lambda\lambda 1661,1666$} & \\
37 & New & 8.5 -- 9.5 & 3 & Same as 13 & \citet{2026arXiv260407076K} \\
38 & \textbf{Archival} & 8.6 -- 9.9 & 6 & Same as 24 & \citet{2025arXiv250615779P} \\
39 & Gz9p3 & 9.3 & 1 & Same as 7 + H$\alpha$ & \citet{2026arXiv260422460B} \\
40 & UNCOVER-26185 & 10.0 & 1 & SED Fitting & \protect\citet{2026arXiv260202323A} \\
41 & \textbf{MACS0647-JD} & 10.2 & 1 & Same as 7 + H$\alpha$ + \textbf{[O\textsc{iii}]$\lambda 4363$} & \citet{2024ApJ...973...81H} \\
42 & JADES & 10.4 -- 13.2 & 4 & SED Fitting & \protect\citet{2023NatAs...7..622C} \\
43 & GHZ-2 & 12.3 & 1 & Same as 13 & \citet{2025NatAs...9..155Z} \\
44 & PAN-z14-1 & 13.5 & 1 & SED Fitting & \citet{2026arXiv260111515D} \\
45 & JADES GS-z14-1 & 13.9 & 1 & SED Fitting & \citet{2025ApJ...992..212W} \\
46 & JADES GS-z14-0 & 14.3 & 1  & SED Fitting & \citet{2025AA...696A..87C}, \\
 &  & &  & & \citet{2025ApJ...988...19S} \\
47 & MoM-z14 & 14.4 & 1 & SED Fitting & \citet{2026OJAp....956033N} \\
\hline
\label{tab:tab2}
\end{tabular}
\end{table*}

\begin{figure}
\centering
\includegraphics[width=\columnwidth]{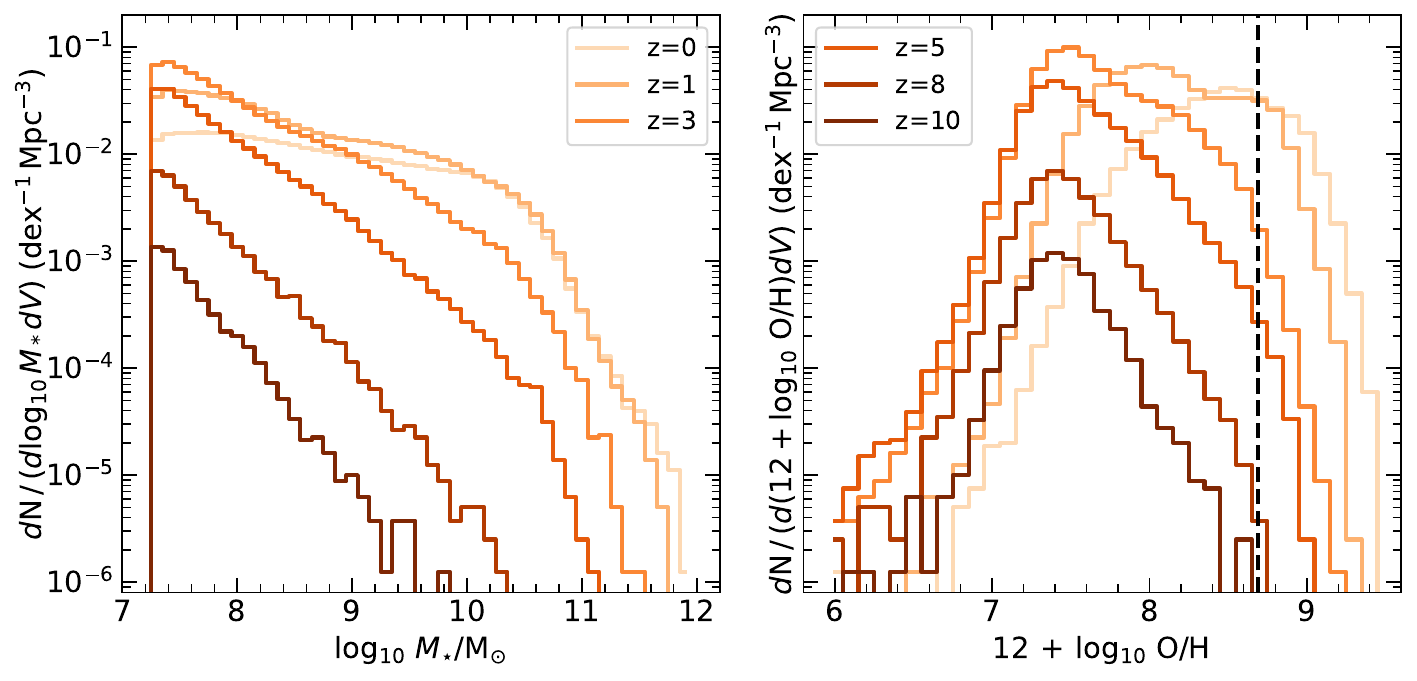}
\caption{\textit{Left panel:} Number density of star-forming COLIBRE galaxies as a function of stellar mass, measured in 0.1 dex mass bins and normalized by the simulation volume, at different redshifts in the L200m6 simulation. Increasingly darker shades of orange correspond to distributions at higher redshifts. The vertical hard cutoff at the low mass end in the distributions arises from only selecting galaxies with $\geq 10$ star particles. \textit{Right panel:} Same as the left panel but for the gas-phase oxygen abundances in $0.1\,\rm{dex}$ bins. The vertical dashed line in the right panel corresponds to Solar metallicity.}
\label{fig:pdf}
\end{figure}

\section{Results}
\label{s:results}
To provide an intuitive view of how galaxy metallicity evolves in the COLIBRE simulations, we begin by examining a single representative galaxy in the L025m5 simulation across cosmic time (see \autoref{fig:visual}). This galaxy evolves into a massive, grand-design spiral by $z=0$, with a stellar mass of $M_{\star} = 1.45\times10^{11}\,\rm{M_{\odot}}$. \autoref{fig:visual} presents 50 ckpc face-on projections (using the stellar angular momentum vector) of its gas-phase metallicity and an RGB composite image of stellar emission in the (observed-frame) near-IR obtained via post processing the galaxy with \texttt{PARTRIDGE} (Huško et al. in preparation). \texttt{PARTRIDGE} generates mock galaxy images by convolving stellar light with different waveband filters, taking into account the effects of 3D dust attenuation and scattering using the live dust model implemented in COLIBRE. We see that the metallicity increases steadily over time, while the spatial distributions of both stars and gas transition from a clumpy and turbulent dust-obscured morphology at high redshift to a smooth, ordered disk by $z=0$, with an inside-out negative metallicity gradient arising due to a metal-rich galaxy centre as compared to the outskirts. The bottom right panel shows the evolution of the stellar mass and gas-phase metallicity of this galaxy as a function of redshift, depicting how this galaxy evolves over cosmic time, and contribute to establishing the MZR at different redshifts. We find that the rise in metallicity is not always monotonic; the metallicity for this galaxy decreases at $z \approx 3$ despite an increase in stellar mass, likely due to an accretion event that dilutes the gas and lowers the metallicity \citep[e.g.,][]{2010ApJ...710L.156R,2018MNRAS.475.1160H,2022MNRAS.509.2720S}.

Next, we examine the normalized density distribution of the stellar mass and metallicity in bins of $0.1\,\rm{dex}$ across all star-forming galaxies in a single COLIBRE simulation (L200m6). \autoref{fig:pdf} shows that this simulation contains galaxies spanning a wide range in stellar masses: $7 \lesssim \log_{10}(M_{\star}/\rm{M_{\odot}}) \lesssim 12$, with metallicities ranging from 0.1 per cent Solar to super-Solar. The sharp cutoff at the low-mass end in $M_{\star}$ reflects our imposed resolution limit: we include only galaxies resolved with at least 10 star particles. At this threshold, we find that the MZR is converged across all three simulation resolutions (see \aref{s:app_boxsize} for convergence with box size at a fixed resolution). Importantly, \autoref{fig:pdf} shows that the COLIBRE simulations contain a statistical sample of resolved galaxies with stellar masses comparable to those observed at all redshifts, particularly during the epoch of reionization and cosmic dawn, making it feasible to assess the simulated and observed MZR together across these early times.

To create the MZR at a fixed redshift, we bin the simulated galaxies in stellar mass bins of width $0.1\,\rm{dex}$, similar to that done in previous works for other cosmological simulations \citep[e.g.,][]{2025MNRAS.536..119G}. Following \citet{chaikin25b} and \citet{2025arXiv250821126S}, we introduce a lognormal scatter in the stellar masses to mimic random errors in stellar mass measurements and account for Eddington bias that results from the low number of galaxies available at the most massive end of the galaxy stellar mass function \citep[][]{2022MNRAS.513..439D}. We adopt the functional form for the lognormal scatter from \citet{2019MNRAS.488.3143B}, modified to use slightly different coefficients by \citet{chaikin25b},
\begin{equation}
\sigma_{M_*} = \mathrm{min}\left(0.1+0.1z, 0.3 \right)\,,
\label{eq:scatter}
\end{equation}
showing how variations in $\sigma_{M_*}$ impact the MZR in \aref{s:app_mstar_random}. Further, at each COLIBRE resolution, we separate mass bins which contain fewer than 20 galaxies from the rest of the sample. In the subsections below, we show the MZR at different redshifts, and compare it with observations and other simulations.

\subsection{Comparison of the COLIBRE MZRs with observations}
\label{s:compare}
\begin{figure*}
\centering
\includegraphics[width=\textwidth]{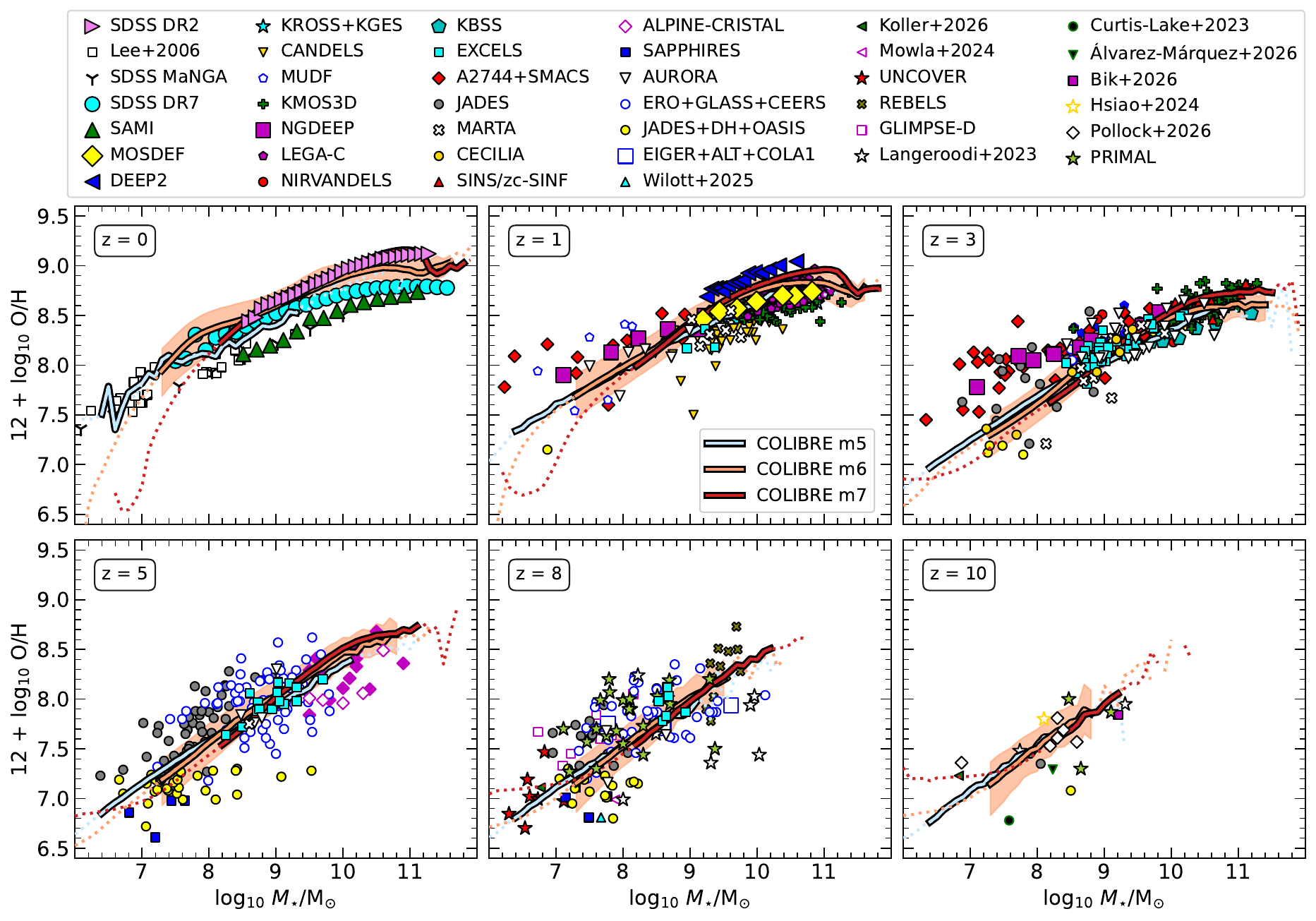}
\caption{Evolution of the median gas-phase mass-metallicity relation (MZR) of star-forming galaxies across cosmic time in the COLIBRE simulations at three different mass resolutions: m5 (light blue), m6 (orange), m7 (red). COLIBRE curves transition from solid to dotted if the number of galaxies drops below 20 in a given stellar mass bin, or if the resolution limit is reached ($<10$ star particles per galaxy). Shaded orange regions indicate the $16^{\rm{th}} - 84^{\rm{th}}$ percentile range for the m6 simulation. Different markers correspond to observational measurements, as listed in the legend and described in \autoref{tab:tab2}. Observational measurements are shown on the redshift panel closest to the overplotted COLIBRE MZRs. Bigger markers represent median values and smaller markers represent individual galaxy measurements. Unfilled markers denote direct $T_{\rm{e}}$ oxygen abundances where at least one auroral line was used, whereas filled markers represent oxygen abundance estimated using strong-line diagnostics. For the MUDF and ALPINE-CRISTAL datasets, direct $T_{\rm{e}}$ abundances are used wherever available, and shown with unfilled markers.}
\label{fig:mzr}
\end{figure*}

We present a summary of all the observational datasets we use for the comparison in \autoref{tab:tab2}.\footnote{The compilation of observational measurements is available at \url{github.com/psharda/colibre_mzr_sharda26}.} \autoref{fig:mzr} shows the median MZRs from the COLIBRE simulations at the three resolutions (m5, m6, m7). The six panels correspond to $z=0,1,3,5,8,10$, respectively. We keep the color scheme consistent throughout this paper: shades of blue, orange and red correspond to COLIBRE simulations at m5, m6 and m7 resolutions, respectively, as in previous works using COLIBRE \citep[e.g.,][]{2025arXiv250821126S,chaikin25b,2026arXiv260403503F}. The COLIBRE MZRs turn dotted at the lowest masses as we reach the resolution limit ($< 10$ star particles), and at the highest masses where we run into low number statistics ($<20$ galaxies in a given mass bin). To provide an estimate of the scatter of the MZRs, we also show the $16^{\rm{th}} - 84^{\rm{th}}$ percentile range around the median for the L200m6 simulation. The scatter around the MZR and correlation with the SFR (so-called fundamental metallicity relation or FMR -- \citealt{2008ApJ...672L.107E,2010MNRAS.408.2115M}) will be discussed in detail in a forthcoming work (Burrafato et al. in preparation). We note that while the MZRs at different resolutions align closely with each other at the low-mass end at all redshifts, larger boxes using coarser resolution show elevated metallicities at the high-mass end at the lowest redshifts. The difference at fixed stellar mass between the m5 and m6 resolutions reaches $0.2\,\rm{dex}$, whereas that between m6 and m7 resolution reaches $0.4\,\rm{dex}$. As we explore later in \autoref{s:agn}, this discrepancy can be attributed to the effects of AGN feedback and the fact that most massive galaxies are not star-forming as per our sSFR threshold.

We also overplot a wide range of observational data available across redshifts in \autoref{fig:mzr}. We overplot the observations on the closest redshift for which we show the COLIBRE MZRs (see subsections below). At the lowest redshifts, we opt to plot median relations from large observational campaigns (see \autoref{tab:tab2} for details), denoted by larger markers. We use filled markers to show measurements that use strong-line diagnostics, and empty markers to show measurements that use at least one auroral line (commonly known as the direct $T_{\rm{e}}$ method, where $T_{\rm{e}}$ represents the electron temperature).

Before we look into the comparison of the data with the simulations, a word of caution is warranted. We note that the observational sample is highly inhomogeneous, using a mix of strong line and auroral line calibrations (often limited by the number of objects available), spanning different aperture sizes, rest-frame wavelengths, S/N of the spectra, and also using somewhat different cuts on common emission line diagnostics such as the \citet{1981PASP...93....5B} BPT diagram. In fact, even in cases where the same line ratios were used, the usage of different calibrations can result in systematic scatter exceeding $0.1\,\rm{dex}$ \citep[e.g.,][]{2025arXiv250810099S}. For some sources, metallicity estimates derived using the same diagnostics can nonetheless vary between studies due to differences in data reduction pipelines \citep[e.g.,][their Section 3.2]{2023ApJ...957...39L}. Furthermore, the uncertainties in stellar mass estimates typically derived from photometry or spectral energy distribution (SED) fitting can also be large, and is sensitive to the assumed IMF, star formation history, dust extinction, etc.  \citep[e.g.,][]{2008MNRAS.388.1595D,2015MNRAS.452.3209R,2015MNRAS.452..235S,2020ApJ...904...33L,2026arXiv260505327G}, especially at high redshifts \citep[e.g.,][]{2024ApJ...963...74W,2025ApJ...978...89H}. As such, the normalization of the MZR carries significant uncertainties. Nonetheless, the slope of the MZR and variations therein carry useful information and is less impacted by the caveats we mention above. In the subsections below, we present the comparison between simulations and observed data at different redshifts which reflect distinct regimes of low and high redshift galaxy evolution.

\subsubsection{Local Universe \texorpdfstring{($z \approx 0$)}{z = 0}}
\label{s:results_z=0}
We begin by examining the $z=0$ panel in \autoref{fig:mzr}. At $z=0$, the MZR in the COLIBRE m7 run has a steeper slope than the m5 and m6 runs, followed by an inflection at $M_* \approx 10^{11.3}\,\rm{M_{\odot}}$. The m5 MZR tends to become noisy at the low-mass end due to the small box we use at $z=0$ ($25\,\rm{Mpc}$), but there is reasonable convergence with resolution at the low-mass end. It is also worth noting that at the highest masses, most galaxies are not considered star-forming as per our sSFR selection criterion. While the normalization of the high-mass end of the MZR can be influenced by our choice of subgrid metal diffusion model, the presence of this turnover is robust to variations in the diffusion prescription \citep[][figure 6]{2026arXiv260400980C}. This inflection reflects that galaxies reach a saturation in metallicity, modulated by AGN feedback that also leads to lower gas fractions in massive galaxies, and has been noted in some other simulations \citep[e.g.,][]{2017MNRAS.467..115D,2017MNRAS.472.3354D,2019MNRAS.484.5587T}, as well as observational data at $z=0$ regardless of aperture effects, star formation activity, N/O ratio and contamination from diffuse ionized gas \citep[][]{2019ApJ...877....6B}. However, the onset of saturation of the MZR can differ between different works, as we discuss below in \autoref{s:results_z=5-8}. We also notice the steeper drop-off of the m6 and m7 resolution MZRs as compared to the m5 run at the low-mass end, reflecting the resolution limits we discussed above.

Since the comparison to observations at $z=0$ has already been discussed in \citet{2025arXiv250821126S}, we keep our analysis brief. We overplot median MZRs from the SDSS survey (DR2) wherein the metallicity was measured using theoretical strong-line calibrations in $3\arcsec$ fibre apertures \citep[][]{2004ApJ...613..898T}. We supplement these with 1) median datapoints from SDSS DR7 from \citet{2020MNRAS.491..944C} using strong-line diagnostics calibrated to a direct $T_{\rm{e}}$-based method that relies on auroral lines to measure the metallicity, 2) best-fitting relation of the measurements for SDSS MaNGA galaxies from \citet{2020AA...634A.107Y} using [O\textsc{ii}], [O\textsc{iii}] and in some cases [N\textsc{ii}] auroral lines to measure the metallicity, and 3) median datapoints from the SAMI galaxy survey \citep[DR3,][]{2022MNRAS.510..320F} where the authors use the same approach as \citet{2020MNRAS.491..944C} to measure the metallicity (within one effective radius) of star-forming galaxies in the SAMI survey \citep[][]{2021MNRAS.505..991C}. There is noticeable systematic scatter among the three SDSS MZRs; the MZR from SDSS DR2 completely overlaps with COLIBRE m7. The offset between the DR2 \citep{2004ApJ...613..898T} and DR7 \citep{2020MNRAS.491..944C} relations is likely driven by the use of different strong-line metallicity calibrations, while the differences between the DR7 and MaNGA \citep{2020AA...634A.107Y} MZRs may arise from selection effects associated with the use of auroral-line measurements in the latter and/or differences in the underlying galaxy SFR distributions. The SAMI galaxies \citep{2022MNRAS.510..320F} lie below the $16^{\rm{th}}-84^{\rm{th}}$ percentile region, but the slope is in good agreement with the COLIBRE m6 MZR.

At the low-mass end ($M_* \lesssim 10^{7.5}\,\rm{M_{\odot}}$, at which comparison with observations was not carried out in \citealt{2025arXiv250821126S}), in addition to the best-fitting relation from \citet{2020AA...634A.107Y}, we also include metallicity data of nearby metal-poor dwarf galaxies compiled by \citet{2006ApJ...647..970L}, where the metallicities were measured using [O\textsc{iii}] auroral line where available, and a combination of [O\textsc{ii}] and [N\textsc{ii}] strong lines otherwise. The COLIBRE m5 simulation displays very good agreement with the MZR for metal-poor local dwarfs ($M_{\star} < 10^{7.5}\,\rm{M_{\odot}}$), where other simulations significantly diverge, and have traditionally failed to reproduce the trends, as we discuss below in \autoref{s:compare_sims}.

\subsubsection{Cosmic Noon \texorpdfstring{($1 \leq z \leq 3$)}{1 <= z <= 3}}
\label{s:results_z=1-3}
Cosmic noon is characterized as the time at which star formation peaked in the Universe, at $z \approx 2$ \citep[][]{2014ARA&A..52..415M}, although the precise redshift at which this occurs remains an active topic of investigation \citep[e.g.,][]{2020MNRAS.494.3828D,2021ApJ...909..165Z,2022ApJ...941...10V,2023MNRAS.518.6142A}. To encapsulate this era, we next examine the MZRs at $z=1$ and $z=3$. On the $z=1$ and $z=3$ panels, we overplot observations that fall within $0.5 \leq z < 2$ and $2 \leq z < 4$, respectively.

We show the resulting median MZRs from COLIBRE at $z=1$ in the top middle panel of \autoref{fig:mzr}. We find good agreement between the three different COLIBRE resolutions at $z=1$. The turnover at the highest masses is only captured in the m6 and m7 runs since the m5 run does not have galaxies with $M_{\star} > 10^{11}\,\rm{M_{\odot}}$ at $z=1$ within its $(50\,\rm{Mpc})^3$ volume. For comparison, we show median datapoints from archival observations and the Keck MOSDEF survey \citep[][]{2015ApJS..218...15K} for which metallicities were estimated by \citet{2026ApJ..1000..109J} using strong-line diagnostics. We also include strong line-based median MZRs from the Keck DEEP2 survey \citep{2011ApJ...730..137Z}, and VLT KMOS KROSS+KGES surveys \citep[][]{2016MNRAS.457.1888S,2020MNRAS.492.1492G}. For the latter, \citet{2021MNRAS.500.4229G} provide measurements for galaxies across $z=0.6-1.8$ in the KROSS and KGES surveys, which we use to construct the median MZR at $z=1$ by taking all galaxies between $0.7 \leq z \leq 1.3$ into account. Green plus markers and red upward triangles denote results from the KMOS3D \citep{2015ApJ...799..209W,2019ApJ...886..124W} and SINS/zc-SINF surveys \citep[][]{2007ApJS..172...70L,2009ApJ...706.1364F}, for which the MZR is presented in \citet{2016ApJ...827...74W} and \citet{2018ApJS..238...21F}. Note that both \citet{2016ApJ...827...74W} and \citet{2018ApJS..238...21F} only present the N2 line ratio in their work; we convert it to an equivalent oxygen abundance by applying the direct $T_{\rm{e}}$-based N2 calibration recommended by \citet{2020MNRAS.491..944C}. We further include strong line-based measurements from the JWST EXCELS survey \citep{2026MNRAS.tmp..397S} and from the HST CANDELS survey that used grism spectroscopy to derive oxygen abundances \citep[][]{2021ApJ...919..143H}, as well as those from the AURORA survey that targeted multiple auroral lines at the sparsely populated low mass end \citep{2025arXiv250810099S,2025arXiv251216989K}. Magenta squares represent median measurements of the low-mass end of the MZR using stacked data from the JWST NIRISS NGDEEP survey \citep[][]{2024ApJ...965L...6B,2026arXiv260520810H}. At the high-mass end, magenta pentagons denote $z \approx 0.7$ data from the VLT VIMOS LEGA-C survey \citep[DR3,][]{2021ApJS..256...44V} presented in \citet{2024ApJ...964...59L}. Red diamonds depict strong-line measurements in the A2744 and SMACS fields using JWST NIRISS grism spectroscopy \citep[][]{2023ApJ...955L..18L}. Blue pentagons denote a mixture of strong-line and direct $T_{\rm{e}}$ measurements from the MUSE Ultra Deep Field \citep[MUDF,][]{2024ApJ...966..228R}. Lastly, we show strong-line-based measurements from \citet{2026arXiv260611345I} who compile and analyze a sample of 50 metal-poor galaxies from the JADES DR4 \citep[][]{2026MNRAS.tmp..935C,2026MNRAS.tmp..886S}, Dark Horse \citep[DH,][]{2026MNRAS.549ag824D} and OASIS JWST programmes; note that only one galaxy in \citet{2026arXiv260611345I} has $ z < 2$, so we plot the rest in subsequent $z\geq 3$ panels.

At $z \approx 1$, the match between the COLIBRE simulations and data from KROSS+KGES, KMOS3D, SINS/zc-SINF, and MOSDEF surveys is very good across the entire mass range covered by the observations. The DEEP2 data from \citet{2011ApJ...730..137Z} are systematically offset from the COLIBRE MZRs by $ \approx 0.05 - 0.1\,\rm{dex}$, but the overall slope of the relation is in good agreement with that in the COLIBRE simulations. This offset may reflect differences in the SFRs of the observed galaxies, consistent with the dependence of metallicity on SFR embodied in the FMR \citep[][]{2010MNRAS.408.2115M}. At the high-mass end, the LEGA-C sample from \citet{2024ApJ...964...59L} shows a saturation in the MZR but not the downturn seen in COLIBRE, although the slope of the LEGA-C MZR is in good agreement with the simulations. This difference is consistent with the LEGA-C sample excluding galaxies that host AGN, given that we show in \autoref{s:agn} that the downturn in COLIBRE is likely driven by AGN feedback. At fixed stellar mass, galaxies in the CANDELS survey are somewhat metal-deficient as compared to COLIBRE and other observational datasets. At the low-mass end ($M_{\star} \sim 10^8\,\rm{M_{\odot}}$), COLIBRE MZRs capture the scatter in AURORA galaxies \citep[][]{2025arXiv250810099S} but show modest ($< 0.2\,\rm{dex}$) offset from the NGDEEP stacks \citep[][]{2026arXiv260520810H}, and large offsets (up to $ 0.7\,\rm{dex}$) from the A2744+SMACS \citep[][]{2023ApJ...955L..18L} and MUDF \citep[][]{2024ApJ...966..228R} dwarf galaxies, respectively. At $M_* \leq 10^7\,\rm{M_{\odot}}$, the latter sample even shows higher metallicities at $z=1$ than at $z=0$. As we discuss below in \autoref{s:snfeedback}, COLIBRE simulations with weaker supernova feedback better reproduce the shallower MZRs from these three datasets. However, too few measurements are currently available in this mass regime to make firm conclusions. 

At $z \approx 3$, in addition to data from the KMOS3D, SINS/zc-SINF, EXCELS, NGDEEP, and AURORA surveys, we include observations from an earlier data release of the JWST JADES survey \citep[][]{2024A&A...690A.288B}, which targeted 146 galaxies at $z \gtrsim 3$ with NIRSpec micro-shutter assembly (MSA), with oxygen abundances derived using $T_{\rm e}$-calibrated strong-line diagnostics \citep{2024AA...684A..75C}. We also overplot abundances derived using multiple auroral lines from the JWST MARTA survey \citep{2025AA...703A.208C} and strong line-based abundances from the JWST CECILIA survey \citep[][]{2025arXiv251200162R} and the VLT/Keck NIRVANDELS survey \citep[][]{2024MNRAS.532.3102S}. Finally, we include median datapoints from the $z \approx 2.3$ Keck MOSFIRE KBSS survey \citep[cyan pentagons in the top right panel,][]{2014ApJ...795..165S}. We opt to use their O3N2 metallicities (based on the \citealt{2004MNRAS.348L..59P} calibration) since the authors report that the O3N2 metallicities show the least bias in cases where direct $T_{\rm{e}}$ abundances are also available. 

The COLIBRE simulations are consistent with a large fraction of the data, particularly with auroral line-based measurements from multiple surveys such as EXCELS, AURORA and MARTA \citep[][]{2026MNRAS.tmp..397S, 2025arXiv250810099S, 2025AA...703A.208C}. At the high-mass end, the predictions are consistent with the trends observed in the KMOS3D and SINS/zc-SINF surveys \citep[][]{2016ApJ...827...74W,2018ApJS..238...21F}. The KBSS median datapoints \citep[][]{2014ApJ...795..165S} are consistent with the COLIBRE m5 and m6 simulations at $M_* \approx 10^{10}\,\rm{M_{\odot}}$ and $10^{11}\,\rm{M_{\odot}}$, respectively \citep[][]{2014ApJ...795..165S}. However, the observed scatter is larger at the low-mass end (more than $1\,\rm{dex}$ at fixed stellar mass). It is also clear that both NGDEEP stacks and A2744+SMACS galaxies follow a much shallower MZR at $z \approx 3$ than COLIBRE. These differences could arise from a combination of systematic uncertainties and the physical nature of low-mass high-redshift galaxies as they go through cycles of bursty star formation, feedback and mergers \citep[e.g.,][]{2016MNRAS.456.2140M,2023MNRAS.526.2665S,2025ApJ...985..126G,2025A&A...697A..88L,2025ApJ...991L...4M}. As we discuss later in \autoref{s:apertures}, the choice of aperture used to measure metallicity can also have a non-negligible impact on the simulated MZRs.

\subsubsection{Epoch of reionization \texorpdfstring{($5 \leq z \leq 8$)}{5 <= z <= 8}}
\label{s:results_z=5-8}
To study the median MZR in the COLIBRE galaxies during and immediately after the epoch of reionization, we look at the results at $z=5$ and $z=8$ in the first two panels of the bottom row of \autoref{fig:mzr}. On the $z=5$ and $z=8$ panels, we overplot observations that fall within $4 \leq z < 6$ and $6 \leq z < 9$, respectively. The high-mass turnover of the MZR in COLIBRE galaxies is only visible for COLIBRE m7 at $z=5$, but disappears otherwise due to the lack of very massive ($M_* \gtrsim 10^{11.3}\,\rm{M_{\odot}}$) star-forming galaxies in the simulation volumes available (see \autoref{fig:pdf}). In addition to the AURORA galaxies from \citet{2025arXiv250810099S}, JADES galaxies from \citet{2024AA...684A..75C} and JADES+DH+OASIS galaxies from \citet{2026arXiv260611345I}, we include measurements from the ALPINE-CRISTAL-JWST survey from \citet{2026ApJS..282...19F} wherein the authors measured oxygen abundances using a combination of strong line (filled magenta diamonds) and $T_{\rm{e}}$-based diagnostics (unfilled magenta diamonds); we use the latter where available. In 4 out of the 5 galaxies where both diagnostics are available, \citet{2026ApJS..282...19F} find that auroral line metallicity estimates are lower than those based on strong lines by $\approx 0.2\,\rm{dex}$. Blue squares represent strong-line measurements of seven galaxies from the JWST SAPPHIRES programme \citep[][]{2025arXiv250503873H}. We also overplot strong-line metallicity estimates from an archival compilation of JWST early release observations \citep[ERO,][]{2022ApJ...936L..14P} together with data from the GLASS \citep{2022ApJ...935..110T} and CEERS surveys \citep{2023ApJ...946L..13F} from \citet{2023ApJS..269...33N}. While COLIBRE galaxies show good agreement with the data at intermediate as well as lower masses, the median MZRs tend to overpredict the metallicity at the high-mass end populated only by the ALPINE-CRISTAL sample. The JADES+DH+OASIS data from \citet{2026arXiv260611345I} seem to follow a very different trend than the rest, with a significantly flatter MZR, but the low-mass, metal-poor end of this sample is in very good agreement with COLIBRE despite the scatter between JADES, SAPPHIRES and JADES+DH+OASIS. In addition to the systematic effects discussed above, this discrepancy may also reflect observational selection biases, as current high-redshift samples of dwarf galaxies are likely skewed toward the brightest -- and potentially more metal-enriched galaxies at fixed stellar mass.

The level of agreement between COLIBRE and observations at $z \approx 8$ can be read off from the bottom middle panel in \autoref{fig:mzr}. Here, we overplot observations from the UNCOVER and REBELS surveys from \citet{2024ApJ...976L..15C} and \citet{2026MNRAS.546f2023R}, respectively. We also overplot data from \citet{2025ApJ...978..136S} and \citet{2025ApJ...985...24C} who present an analysis of galaxies in the PRIMAL survey together with archival data, and \citet{2023ApJ...957...39L} who compile a sample of 11 galaxies at $z\approx 8$ from the literature (including two new sources). The metallicity estimates are derived from a mix of strong-line and direct $T_{\rm{e}}$ methods, and also via IR emission lines in some cases. Lastly, we plot results for three galaxies resolved with NIRSpec-IFU from \citet{2026arXiv260407076K}, the gravitationally-lensed Lyman-$\alpha$ emitter CANUCS-A370-z8-LAE from \citet{2025ApJ...988...26W}, and direct $T_{\rm{e}}$ abundance measurement for the $z=8.3$ Firefly Sparkle galaxy from \citet{2024Natur.636..332M}. We also include stacked measurements based on the [O\textsc{iii}] auroral line in the EIGER, ALT and COLA1 surveys from \citet{2026AA...706A.165K}. This panel underscores the importance of combining simulation volumes for such comparisons: the high-resolution L100m5 COLIBRE box provides a statistically significant sample of low-mass galaxies at $z=8$, while the larger L400m7 box captures a substantial population of the most massive systems observed at this epoch, together enabling a robust comparison with observations across the full stellar mass range. 

If the UNCOVER sample from \citet{2024ApJ...976L..15C} were excluded, it would seem that COLIBRE underpredicts median oxygen abundances at this epoch. However, the predictions agree very well with \citet{2024ApJ...976L..15C}, which highlights the importance of measuring metallicities in very low-mass systems that place critical constraints on the slope of the MZR. At intermediate masses, COLIBRE predictions are in good agreement with direct $T_{\rm{e}}$ measurements from the ERO+GLASS+CEERS and EXCELS programmes \citep[][]{2023ApJS..269...33N,2026MNRAS.tmp..397S}. We note that for the most massive galaxies, such as those in the REBELS survey \citep[][]{2026MNRAS.546f2023R}, strong-line-based abundance estimates may be systematically overestimated, similar to what has been reported for the ALPINE-CRISTAL sample at $z=5$ \citep[][]{2026ApJS..282...19F}. On the other hand, it could also be the case that stellar masses for evolved galaxies are underestimated due to the outshining effect of younger stellar populations over older ones \citep[e.g.,][]{2018MNRAS.476.1532S,2023ApJ...948..126G,2024ApJ...961...73N}, or due to spatially varying dust attenuation that can obscure a considerable fraction of the stellar population \citep[][]{2026arXiv260413408L}. If such systematic offsets at the high-mass end are common (e.g., due to different underlying star formation histories), they would further improve the agreement with the COLIBRE MZRs during the EoR (see also, \autoref{s:apertures} for the effects of aperture sizes). Nevertheless, some galaxies, particularly from \citet{2023ApJ...957...39L}, lie more than $1\,\rm{dex}$ away from the COLIBRE median relation and other observations, possibly highlighting the large scatter due to a mix of systematic biases and the inherent nature of EoR galaxies undergoing rapid evolutionary stages that can boost or reduce galaxy metallicity and drive it out of equilibrium \citep[][]{2021MNRAS.502.5935S,2026AA...706A.165K}.

\subsubsection{Cosmic dawn \texorpdfstring{($z \approx 10$)}{z ~ 10}}
\label{s:results_z=10}
JWST has ushered in observational studies of the era of galaxy formation at cosmic dawn, with several spectroscopically confirmed galaxies at $z \gtrsim 10$ \citep[e.g.,][]{2023A&A...677A..88B,2023NatAs...7..622C,2024Natur.633..318C,2025NatAs...9..155Z,2025ApJ...988...19S}, some of which also have metallicity estimates based on a variety of strong-line diagnostics, as well as SED fitting (except for MACS0647-JD which was detected in [O\textsc{iii}]$\lambda 4363$ -- \citealt{2024ApJ...973...81H}). While the current sample size is not large, it is nonetheless interesting to compare COLIBRE predictions with available measurements at such early epochs when galaxies are only starting to build up their mass and metals. We show the COLIBRE MZRs at $z=10$ in the bottom right panel of \autoref{fig:mzr}, overlaid with measurements at $9 \leq z \leq 10.5$ from \citet{2023NatAs...7..622C,2024ApJ...973...81H,2025arXiv250615779P,2025ApJ...978..136S,2026arXiv260422460B} and \citet{2026arXiv260202323A}. The first key observation is that the COLIBRE volumes include a statistically meaningful sample of galaxies with stellar masses comparable to those inferred from JWST observations at these redshifts, demonstrating that the presence of bright galaxies in the early Universe can be naturally reproduced in these simulations \citep[see also,][]{chaikin25b}. Secondly, COLIBRE simulations show that the MZR is already in place at these redshifts. The comparison clearly demonstrates that the median trend in the observations is well reproduced in the COLIBRE simulations at $z \approx 10$, even though the observed sample size is quite small. As in previous panels, the scatter in the data increases at the low-mass end, likely due to reasons mentioned above. 

\begin{figure}
\centering
\includegraphics[width=\columnwidth]{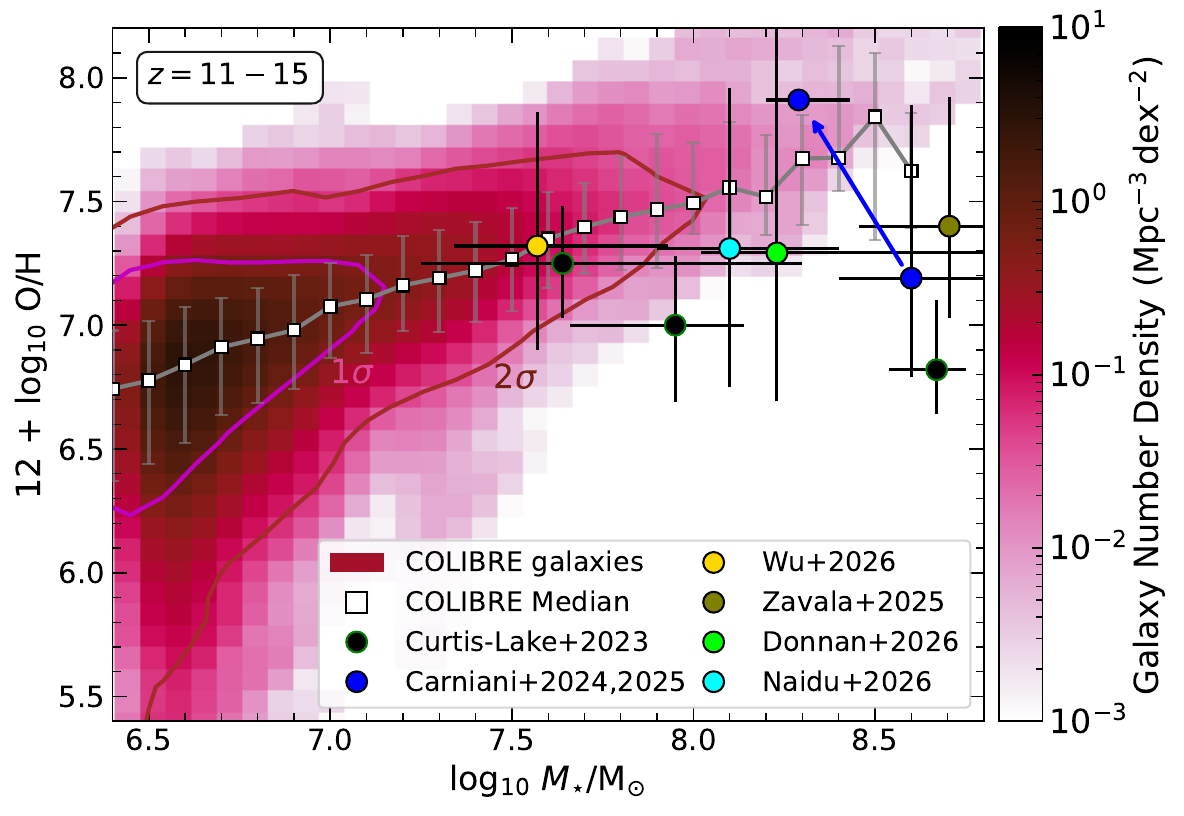}
\caption{Gas-phase metallicity as a function of stellar mass for all resolved $z = 11-15$ star-forming galaxies in the COLIBRE simulations. The smoothed, simulation volume-weighted 2D histogram includes star-forming galaxies at these redshifts at all three resolutions (m5, m6, m7). The contours depict the $1\sigma$ and $2\sigma$ regions for the galaxy number density distribution per dex stellar mass per dex metallicity. White squares denote the volume-weighted median COLIBRE MZR, and grey errors depict the the $16^{\rm{th}} - 84^{\rm{th}}$ percentile range. The median is only plotted for bins with at least 20 galaxies. Overlaid are observational measurements of $z \geq 11$ galaxies from \protect\citet{2023NatAs...7..622C} and \protect\citet{2025ApJ...992..212W}, as well as measurements for galaxies GS-z14-0 from JWST \protect\citep[][]{2024Natur.633..318C} and subsequent JWST+ALMA analyses \protect\citep{2025ApJ...988...19S,2025AA...696A..87C}, GHZ-2 from \protect\citet{2025NatAs...9..155Z}, PAN-z14-1 from \protect\citet{2026arXiv260111515D}, and MoM-z14 from \protect\citet{2026OJAp....956033N}.}
\label{fig:mzr_contours}
\end{figure}

\begin{figure*}
\centering
\includegraphics[width=\textwidth]{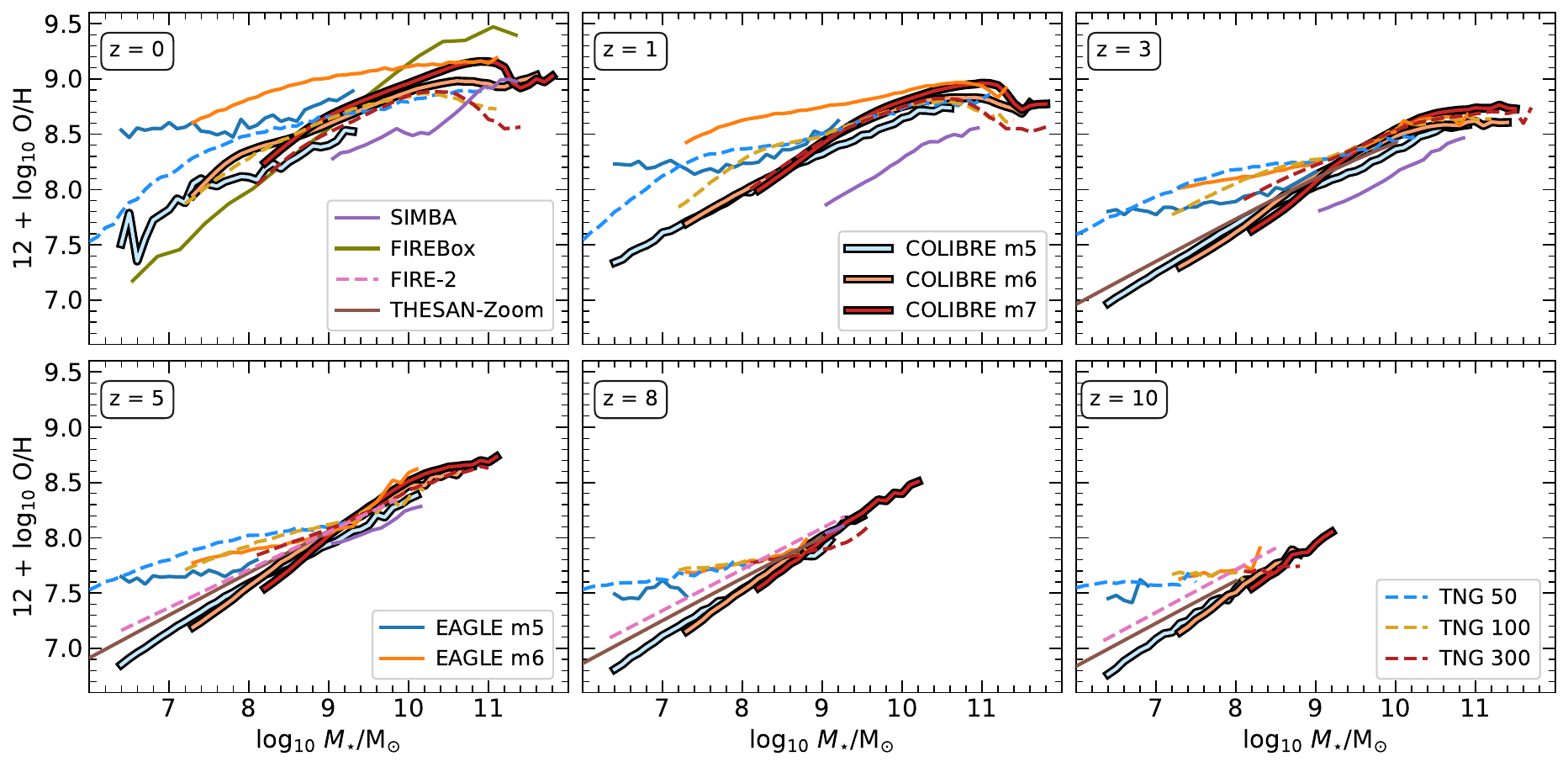}
\caption{Same as \autoref{fig:mzr} but now comparing MZR predictions from COLIBRE with other cosmological simulations. Overplotted are MZRs from EAGLE at m5 and m6 resolutions, TNG 50, 100 and 300 (see \autoref{s:compare_sims} for details), SIMBA at $z=0-5$ \protect\citep[adopted from][]{2025MNRAS.536..119G}, FIREBox at $z=0$ \protect\citep[][]{2023MNRAS.522.3831F}, as well as the best-fit relations from FIRE-2 and THESAN-Zoom at $z \geq 5$ and $z \geq 3$ respectively \protect\citep{2024ApJ...967L..41M,2026MNRAS.tmp...20M}. Notice that the vertical axes limits differ from those in \autoref{fig:mzr}. Apart from COLIBRE, only EAGLE and TNG simulations contain galaxies that satisfy all the selection criteria at all the redshifts shown.}
\label{fig:mzr_sims}
\end{figure*}

\subsubsection{Evolution of the MZR at \texorpdfstring{$z > 10$}{z > 10}}
\label{s:cosmicdawn}
Finally, in \autoref{fig:mzr_contours}, we look at the predictions from COLIBRE for the MZR at the highest redshifts ($z=11-15$), an epoch where at least some galaxies detected in JWST photometry have been spectroscopically confirmed \citep[e.g.,][]{2023NatAs...7..622C,2024Natur.633..318C,2025ApJ...988...19S,2025NatAs...9..155Z}. Instead of only showing the simulation volume-weighted median MZRs at individual redshifts, we opt to collate all the available data from COLIBRE L100m5, L200m6 and L400m7 boxes and also show the simulation volume-weighted, normalized density distribution of galaxies in the mass-metallicity plane. Following the approach in previous sections, we only include galaxies that satisfy the sSFR threshold we introduced in \autoref{s:metallicity_estimation}, are resolved by at least 10 star particles. We only plot the median in $0.1\,\rm{dex}$ stellar mass bins with at least 20 galaxies.

We find that the vast majority of COLIBRE galaxies occupy a low-mass, low-metallicity regime at these epochs. The median MZR traces the underlying galaxy number density quite well except at the highest masses. We also see that observations are beginning to probe the most massive galaxies at these redshifts, as evidenced by the concentration of observational data points toward the upper-right region of \autoref{fig:mzr_contours}. Out of the eight $z \geq 11$ galaxies for which metallicity estimates are currently available \citep[][]{2023NatAs...7..622C,2024Natur.633..318C,2025AA...696A..87C,2025ApJ...992..212W,2025NatAs...9..155Z,2026arXiv260111515D,2026OJAp....956033N}, five galaxies (GS-z12-0, GS-z13-0, GS-z14-1, PAN-z14-1 and MoM-z14) lie within the $2\sigma$ density contours of the simulated galaxy population (within errors), whereas one (GS-z11-0, $M_{\star} \approx 10^{8.7}\rm{M_{\odot}}$) exhibits some offset from the distribution. The metallicity estimates for most galaxies are consistent with the median trends, except for GS-z11-0 and GS-z13-0 from \citet{2023NatAs...7..622C}. However, systematic issues identified above limit a detailed comparison of simulations with observations. In fact, significant differences in inferred metallicities and stellar masses can arise when only JWST data are used for SED fitting at these epochs, compared to analyses that incorporate both JWST and ALMA observations. We illustrate this in \autoref{fig:mzr_contours} with the unfilled blue markers corresponding to galaxy GS-z14-0, where the arrow highlights the shift in both stellar mass and metallicity once rest-frame sub-millimeter data are included \citep[][]{2025ApJ...988...19S,2025AA...696A..87C}. This example underscores the need for multi-wavelength constraints to accurately estimate galaxy properties at these epochs.

As a side note, it is worth mentioning that variations in the IMF at $z > 10$, as expected from star formation theory \citep[e.g.,][]{2022MNRAS.509.1959S,2022MNRAS.514.4639C,2025MNRAS.537..752B} and often invoked to explain current observations \citep[e.g.,][]{2022ApJ...938L..10I,2024MNRAS.534..523C,2024MNRAS.527.5929Y,2025ApJ...980...10J,2025A&A...694A.254H,2026MNRAS.546f2267C} likely play an important role in setting the MZR at these redshifts. Whether IMF variations can have a larger impact on the comparison between simulations and observations than the other effects we discuss below in \autoref{s:discussion}, however, remains an open question that we plan to address in a companion paper (Durrant et al. in preparation).

\subsection{Comparison of COLIBRE MZRs with previous simulations}
\label{s:compare_sims}
In this section, we compare COLIBRE MZRs with other large-scale cosmological and zoom-in simulations. We first provide a brief overview of the data we compile in \autoref{s:compare_sims_compile}, followed by the analysis in \autoref{s:compare_sims_analysis}. We scale the metallicity to $12 + \log_{10}\rm{O/H} = 8.69$ \citep[][]{2009ARA&A..47..481A} wherever needed to be at par with COLIBRE.

\subsubsection{Compilation of simulation data}
\label{s:compare_sims_compile}
We use publicly available data for the EAGLE and TNG simulations at different resolutions. For EAGLE, we consider the recalibrated $25\,\rm{Mpc}$ box at m5 resolution and the $100\,\rm{Mpc}$ box at m6 resolution \citep[][]{2015MNRAS.446..521S,2015MNRAS.450.1937C}. For TNG, we use the $51.7\,\rm{Mpc}$, $110.7\,\rm{Mpc}$, and $302.6\,\rm{Mpc}$ boxes at m5, m6, and m7 resolutions, respectively (commonly referred to as TNG50, TNG100, and TNG300; \citealt{2019MNRAS.490.3196P,2019ComAC...6....2N}). These resolution labels refer to baryons; because COLIBRE employs four times as many dark matter particles at a given resolution, an m$_X$ run in EAGLE or TNG approximately corresponds to m$_{X+1}$ in COLIBRE in terms of dark matter resolution.

To enable a consistent comparison, we adopt similar definitions of stellar mass and gas-phase metallicity. For EAGLE, we use stellar masses measured within a $50\,\rm{kpc}$ aperture, and mass-weighted metallicities computed from star-forming gas bound to each subhalo. For TNG, stellar masses are obtained from all star particles bound to the subhalo, while mass-weighted metallicities are calculated from all star-forming gas cells bound to the subhalo. We further impose the same sSFR threshold, stellar mass binning, resolution limits, and minimum galaxy counts per bin (at least 20 galaxies) for EAGLE and TNG as we used above for COLIBRE. In addition, we apply the redshift-dependent lognormal scatter in \autoref{eq:scatter} above to EAGLE and TNG stellar masses to account for random errors in stellar mass measurements. The MZRs we obtain for EAGLE and TNG are in good agreement with the ones previously published from these simulations using somewhat different criteria \citep[][]{2017MNRAS.472.3354D,2019MNRAS.484.5587T}, demonstrating the robustness of the inferred metallicities. We do not include the $z \geq 3$ MZRs from the recently introduced LUMINA simulations \citep[][figure 13]{2026arXiv260515310Z}, which are comparable to COLIBRE m6 in resolution, because the selection criteria used to construct the MZRs are not clearly specified. Nevertheless, we note that the slope of the LUMINA MZRs are consistent with TNG100.

We also include results from the SIMBA cosmological simulations, using the compilation of \citet{2025MNRAS.536..119G} for the largest ($100\,\rm{Mpc}$) SIMBA box \citep[][]{2019MNRAS.486.2827D}. The baryonic resolution of SIMBA is comparable to COLIBRE m7, but \citet{2025MNRAS.536..119G} only include galaxies resolved with at least 100 star particles and 500 gas particles, which is stricter than our resolution criterion and can potentially induce a selection bias when compared with other simulations. Similar to our approach, \citet{2025MNRAS.536..119G} select star-forming gas to compute mass-weighted metallicities. Their sSFR threshold differs slightly, excluding galaxies more than 0.5 dex below the star-forming main sequence (SFMS) at each redshift; however, they note that the resulting MZR is largely insensitive to this choice. At $z=0$, we further include the median MZR from the FIREBox simulations \citep[][]{2023MNRAS.522.3831F}, which have a box size of $22\,\rm{Mpc}$ and a baryonic particle mass of $6\times10^4\,\rm{M_{\odot}}$. \citet{2023MNRAS.522.3831F} measure the oxygen abundances within a $3\,\rm{kpc}$ aperture, consistent with our COLIBRE measurements, which they further reduce by 0.12 dex to approximately account for dust depletion, following \citet{2010ApJ...724..791P}. 

At higher redshifts ($z \geq 5$), bursty star formation, a rapidly evolving ionizing background, and denser ISM conditions could give rise to distinct MZRs. To explore this possibility, we also include results from cosmological zoom-in simulations that often offer improved mass and spatial resolution, although at the cost of smaller sample sizes and not being run all the way to $z=0$. A number of such zoom-in studies are now available (e.g., Renaissance, FirstLight, FIRE-2, FLARES, SERRA, MEGATRON, THESAN-Zoom; \citealt{2015ApJ...807L..12O,2017MNRAS.470.2791C,2018MNRAS.480..800H,2021MNRAS.500.2127L,2025A&A...699A...6P,2025arXiv251005201K,2025OJAp....8E.153K}). Out of these, we include the FIRE-2 and THESAN-Zoom simulations \citep{2018MNRAS.480..800H,2018MNRAS.478.1694M,2025OJAp....8E.153K}, for which median trends or best-fitting relations are reported in the literature.

The FIRE-2 zoom-in simulations are drawn from a parent dark matter-only run and span $z = 5-9$, with baryonic masses ranging from $100\,\rm{M_{\odot}}$ to $7000\,\rm{M_{\odot}}$ \citep[][]{2018MNRAS.480..800H}. The median MZRs are presented in \citet{2024ApJ...967L..41M}, who estimate the stellar masses and mass-weighted metallicities using all particles within 20 per cent of the virial radius of each subhalo, finding negligible sensitivity to alternative aperture definitions or temperature cuts. When plotting the FIRE-2 MZRs, we only include stellar mass bins where at least five galaxies are available at each redshift \citep[following][figure 1]{2024ApJ...967L..41M}.

The THESAN-Zoom simulations \citep[][]{2025OJAp....8E.153K} are similarly drawn from the parent THESAN runs \citep[][]{2022MNRAS.511.4005K} and re-simulated at higher resolution (baryonic particle mass $142-9090\,\rm{M_{\odot}}$). In these simulations, stellar masses are computed from all bound particles within the virial radius, while mass-weighted metallicities are derived from gas within twice the stellar half-mass radius. We include the best-fit MZR relations spanning $3 < z < 12$ and $6 \leq \log_{10}\,M_{\star}/\rm{M_{\odot}} \leq 11$ from \citet{2024ApJ...967L..41M}, but limit the MZR to the maximum stellar masses available at each redshift \citep[following][figure 4b]{2025OJAp....8E.153K}.

Unlike COLIBRE, neither EAGLE nor TNG or SIMBA explicitly model the cold phase of the ISM. Further, EAGLE, TNG, SIMBA, FIREBox and FIRE-2 do not include a live dust model or the effects of dust depletion. Unlike COLIBRE and EAGLE, the metal mass loading factor in the TNG simulations was tuned to match the MZR.
Overall, this compilation of simulations -- while somewhat heterogeneous in the details of how metallicities and stellar masses are measured -- provides an informative benchmark against which to compare the COLIBRE results.

\subsubsection{Analysis of MZRs from simulations}
\label{s:compare_sims_analysis}
We overplot the MZRs from all the simulations described above in \autoref{fig:mzr_sims}. The COLIBRE MZRs are the same as in \autoref{fig:mzr}, but we do not show the $16^{\rm{th}}-84^{\rm{th}}$ percentile range for clarity. We find that the EAGLE m5 and m6 MZRs probe similar ranges in stellar mass as COLIBRE m5 and m6 at $z=0$, but this is not the case at high-$z$, where COLIBRE MZRs have a significantly larger dynamic range than EAGLE thanks to their large volumes. In contrast, the TNG MZRs reach lower stellar masses than COLIBRE at all redshifts, but fail to sample the most massive galaxies at $z \geq 8$. A similar limitation applies to the FIRE-2 and THESAN zoom simulations, although these were not evolved down to $z=0$, preventing a direct comparison at low redshifts. \citet{2023MNRAS.522.3831F} only study the $z=0$ MZR in the FIREBox simulations, which probes the same stellar mass range as COLIBRE. Owing to its low resolution and small volume, SIMBA primarily samples the high-mass end of the MZR and as per the resolution criterion used in \citet{2025MNRAS.536..119G}, does not contain any massive, resolved galaxies at $z \gtrsim 8$.

\begin{figure*}
\centering
\includegraphics[width=0.85\textwidth]{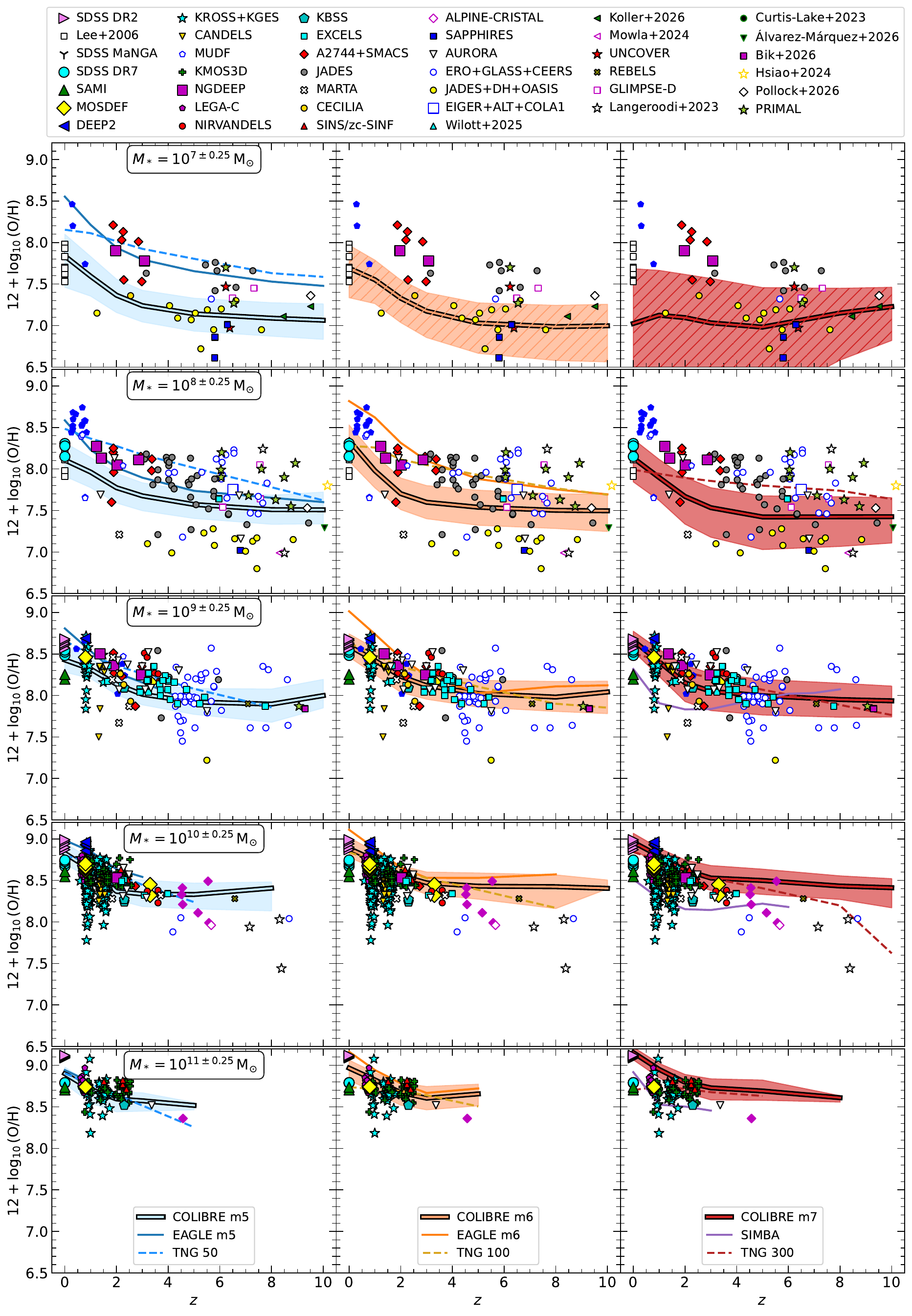}
\caption{Evolution of the MZR with redshift in galaxies within different stellar mass bins per row ($M_{\star} \approx 10^7,\,10^8,\,10^9,\,10^{10},\,10^{11}\,\rm{M_{\odot}}$), separated into three columns corresponding to different COLIBRE resolutions (m5, m6, m7). Shaded bands denote the $16^{\rm{th}} - 84^{\rm{th}}$ percentile ranges; the shaded bands are hatched for COLIBRE m6 and m7 in the top row to signify that they are affected by the resolution limit. Observational measurements are the same as in \autoref{fig:mzr} and follow the same style for strong-line and direct $T_{\rm{e}}$-based measurements, as well as marker size for individual and median datapoints. Results from other cosmological simulations, EAGLE and TNG (see \autoref{s:compare_sims_compile}), and SIMBA \protect\citep{2025MNRAS.536..119G}, are overplotted only in the columns with comparable resolution to the corresponding COLIBRE run.}
\label{fig:redshift_evolution_all}
\end{figure*}

We note that the COLIBRE simulations display excellent convergence across m5, m6, and m7 resolutions at all redshifts (see also, \aref{s:app_boxsize} for convergence with box size at fixed resolution). By contrast, both the EAGLE and TNG suites show substantial resolution-dependent variation, with convergence only established in these simulations at $z \geq 3$ and $z \geq 5$, respectively.\footnote{TNG is not re-calibrated at different resolutions.} We see that both the EAGLE m5 and m6 MZRs exhibit systematically higher metallicites and shallower slopes than COLIBRE m5 and m6, respectively, at all redshifts. The offset in the metallicity reaches as much as 1 dex at the low-mass end, placing the low-mass end of the EAGLE MZRs far above the observations we show in \autoref{fig:mzr}. On the other hand, the TNG 100 and TNG 300 MZRs match the COLIBRE m6 and m7 MZRs very well at low $z$, with TNG 300 showcasing a similar turnover at the massive end to that of COLIBRE m7, although the location and strength of the turnover are different, generating up to 0.4 dex differences in the metallicity at low redshifts ($z \lesssim 1$). At high-$z$, TNG 300 and COLIBRE m7 almost completely overlap, although TNG 300 loses statistics and hence cannot probe a large range in stellar mass at the earliest epochs ($z \gtrsim 8$). Compared to COLIBRE and observed data, the MZRs from EAGLE and TNG at equivalent resolutions become systematically flat and shallow at low masses, particularly at high redshifts.

The MZRs from SIMBA show a similar slope as COLIBRE, but are systematically offset to lower metallicities by $\approx 0.4\,\rm{dex}$, and do not show a clear saturation or turnover. The top left panel of \autoref{fig:mzr_sims} demonstrates that the $z=0$ MZR in FIREBox is much steeper as compared to COLIBRE, EAGLE, TNG and obervations. The FIREBox MZR also exhibits an inflection at the high-mass end, but the turnover is less strong, leading to an overestimation of metallicity for the most massive galaxies, likely due to the exclusion of AGN feedback \citep[][]{2023MNRAS.522.3831F}. We also see that FIRE-2 and THESAN-Zoom simulations deviate by less than $0.2\,\rm{dex}$ from COLIBRE across $z=5-10$.

\begin{figure}
\centering
\includegraphics[width=\columnwidth]{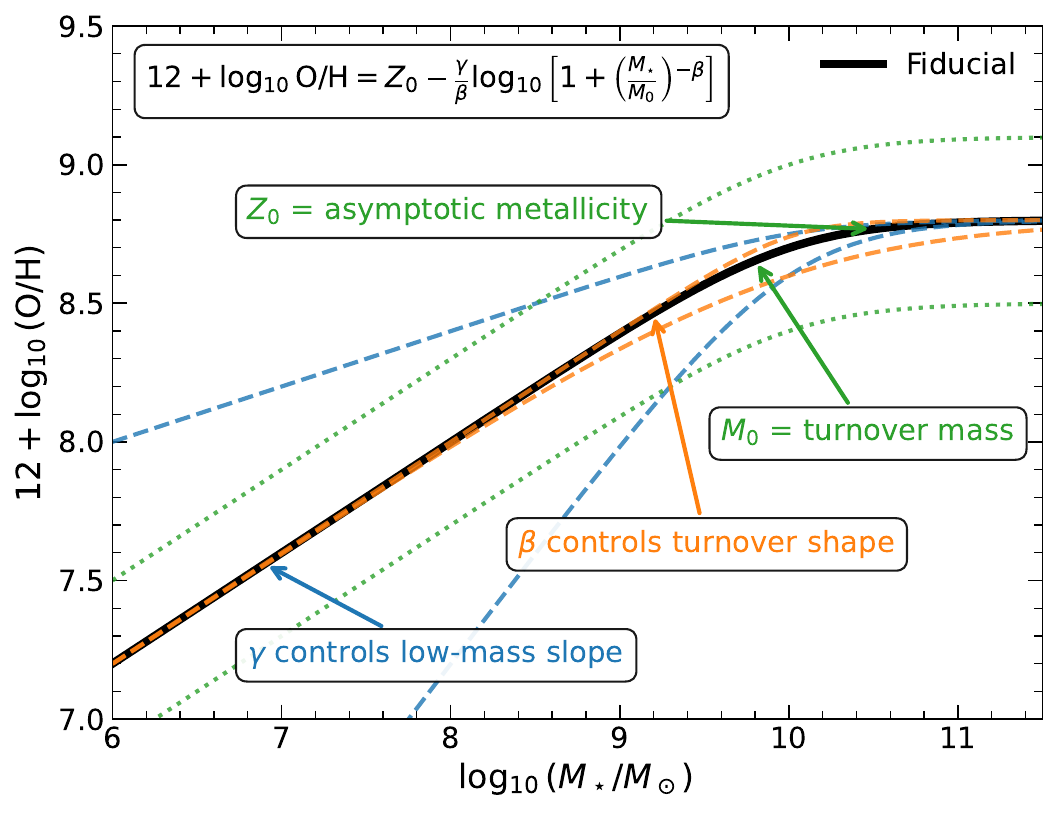}
\caption{Schematic explaining how the free parameters in the broken power-law model (\autoref{eq:fits}) proposed by \protect\citet{2020MNRAS.491..944C} determine key aspects of the MZR.}
\label{fig:schematic}
\end{figure}

\subsection{Evolution of the MZR with redshift}
\label{s:redshift_evolution}
We now investigate variations in the metallicity with redshift at fixed stellar mass, to examine how galaxies with similar stellar mass build up their metal content across cosmic time. In \autoref{fig:redshift_evolution_all}, we show the median redshift evolution of metallicity in COLIBRE galaxies of $M_{\star} = 10^{7},\,10^{8},\,10^9,\,10^{10}$ and $10^{11}\,\rm{M_{\odot}}$, in stellar mass bins of width $0.25\,\rm{dex}$. COLIBRE data show very good convergence between with resolution across all the mass bins, even at $M_* \approx 10^7\,\rm{M_{\odot}}$ where the m6 and m7 simulations (shown with hatches in the top middle and right panels) fall at or below our nominal resolution threshold.


\autoref{fig:redshift_evolution_all} shows that both COLIBRE m5 and m6, are able to not only reproduce the metal-poor end of the MZR at $z=0$ (where the observational points are adopted from \citealt{2006ApJ...647..970L}), but also at high-redshifts. However, COLIBRE significantly underpredicts the metallicity for NGDEEP stacks \citep[][]{2026arXiv260520810H}, MUDF \citep[][]{2024ApJ...966..228R}, and A2744+SMACS galaxies \citep[][]{2023ApJ...955L..18L} at these stellar masses, which show similar or even higher levels of metal enrichment as the $z=0$ local dwarfs \citep[][]{2006ApJ...647..970L}. In comparison, both EAGLE and TNG systematically overpredict the metallicities in this mass range, particularly at $z > 0$, but they can reproduce the metallicity for certain datasets such as NGDEEP, A2744+SMACS, MUDF and JADES. As for COLIBRE m7, the resolution of EAGLE m6 and TNG 300 is likely insufficient to draw a meaningful comparison at $M_* \approx 10^7\,\rm{M_{\odot}}$, so we omit them from the top row in \autoref{fig:redshift_evolution_all}. COLIBRE also reproduces the observed metallicities across all redshifts at $M_\star \approx 10^8\,\rm{M_\odot}$, but the scatter in the data at high $z$ exceeds 1 dex. If the recent JADES+DH+OASIS data from \citet{2026arXiv260611345I} were excluded, it would seem that COLIBRE predicts lower metallicities at intermediate redshifts ($z \approx 1 - 8$) than that observed whereas both EAGLE and TNG are consistent with the observations, which highlights the need for larger samples and also showcases the rapidly evolving metallicity frontier thanks to JWST. At $M_\star \approx 10^9\,\rm{M_\odot}$, the observed evolution of metallicity with redshift is in very good agreement with the COLIBRE predictions across the full redshift range. Compared to COLIBRE, both EAGLE and TNG predict a steeper decline in metallicity with increasing redshift at fixed stellar mass, although their predictions remain broadly consistent with the observational constraints. 

The bottom two rows of \autoref{fig:redshift_evolution_all} show the redshift evolution of the metallicity in massive galaxies, with $M_{\star} = 10^{10\pm0.25}\,\rm{M_{\odot}}$ and $10^{11\pm0.25}\,\rm{M_{\odot}}$, respectively. At $M_* = 10^{10\pm0.25}\,\rm{M_{\odot}}$, we find that COLIBRE m5 is in good agreement with the data up to $z \approx 7$, whereas both m6 and m7 tend to overpredict the median metallicity beyond $z > 5$. A similar trend is seen in EAGLE, which remains consistent with the observations at m5 resolution but increasingly overpredicts metallicities at m6 resolution. In contrast, TNG exhibits somewhat better agreement at high redshift; in particular, TNG 300 shows a steeper decline in metallicity beyond $z > 6$, bringing it into closer agreement with the limited observational constraints currently available.
SIMBA is distinct in predicting a turnover followed by a shallow increase in metallicity at $z \approx 4-6$, in contrast to the more or less monotonic evolution seen in the other simulations. SIMBA systematically underpredicts the median metallicities, but the trend cannot be ruled out with the current data.

\begin{figure*}
\centering
\includegraphics[width=\textwidth]{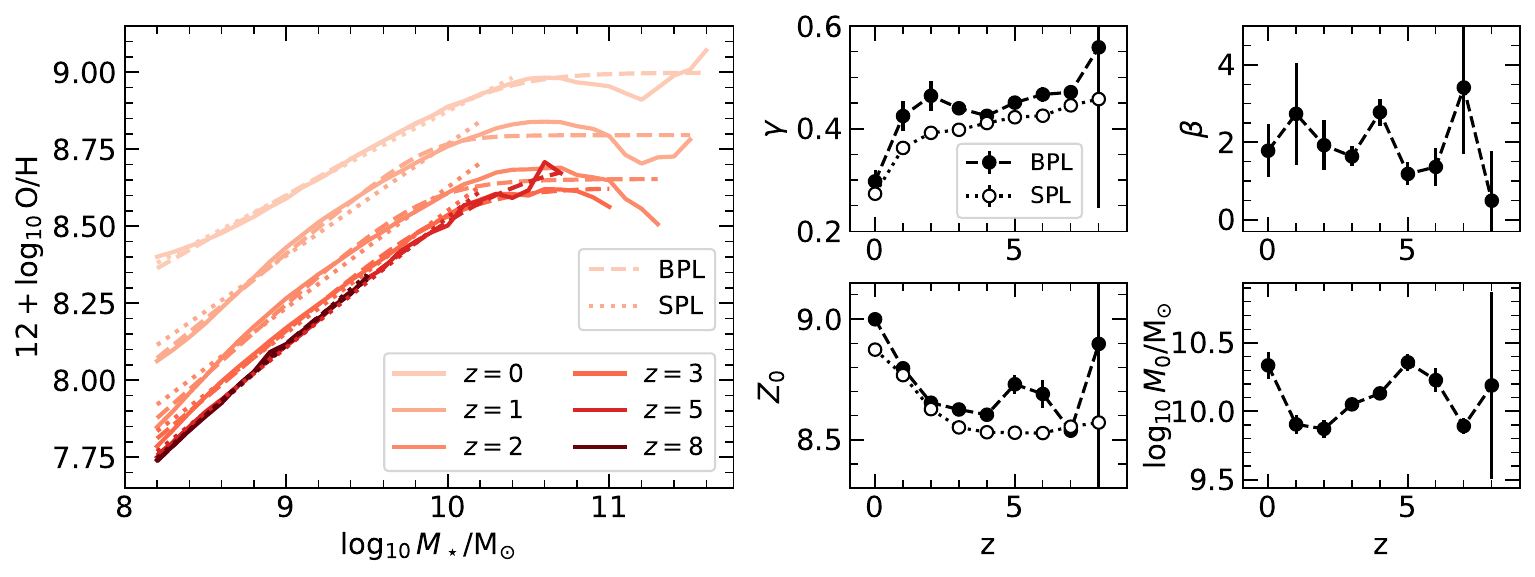}
\caption{\textit{Left panel:} Evolution of the median MZR in the COLIBRE L200m6 simulation at different redshifts (solid). Dashed and dotted lines represent broken power-law (BPL, \autoref{eq:fits}) and single power-law (SPL, \autoref{eq:single-powerlaw}) fits to the MZRs, respectively. \textit{Right panels:} Redshift evolution of the best-fit parameters (slope $\gamma$, inflection $\beta$, saturation metallicity $Z_0$ and turnover mass $M_0$) when the median MZRs in the left panel are fit with the BPL model (dashed) and the SPL model (dotted).}
\label{fig:mzr_fits}
\end{figure*}

At the highest stellar masses, at m5 and m6 resolutions, COLIBRE, EAGLE and TNG are all largely consistent with the data up to $z \approx 3$, but tend to overpredict the metallicity at higher redshifts. At m7 resolution, SIMBA predicts systematically lower metallicities than both COLIBRE and TNG 300. However, the simulation contains too few galaxies with $M_* \approx 10^{11}\,\rm{M_{\odot}}$ beyond $z \approx 3$ to enable a meaningful comparison with the available observations at higher redshifts. The observational constraints themselves are also extremely sparse in this regime, further limiting our ability to distinguish between the models. Note that the results from COLIBRE are also sensitive to AGN feedback and the aperture sizes we use within this mass range, aspects we discuss below in \autoref{s:agn} and \autoref{s:apertures}.

\subsection{Functional fits to median MZRs}
\label{s:fits}
To quantify the shape of the MZR, a few different functional forms have been proposed in the literature \citep[e.g.,][]{2011arXiv1112.3300M,2014ApJ...791..130Z,2020MNRAS.491..944C,2021ApJ...914...19S,2023ApJ...942...24S,2024ApJ...967L..41M,2026MNRAS.tmp...20M}. The basic idea behind these parametric functions is to capture variations in the slope and turnover of the relation, and to explore whether the data can be described by a single function across a wide range in stellar mass. We fit the median COLIBRE MZRs to the broken power-law (BPL) function proposed by \citet{2020MNRAS.491..944C}:

\begin{equation}
12 + \log_{10}\mathrm{O/H} = Z_0 - \frac{\gamma}{\beta} \log_{10}\left[1 + \left(\frac{M_{\star}}{M_0}\right)^{-\beta}\right]\,,
\label{eq:fits}
\end{equation}
where $\gamma$ is the power-law slope at the low-mass end, $Z_0$ and $M_0$ correspond to the asymptotic metallicity and turnover stellar mass respectively, and $\beta$ controls the width of the transition from power-law to constant metallicity. The schematic in \autoref{fig:schematic} illustrates the effect of the different parameters. 

We use the \texttt{LMFIT} package in \texttt{python} to fit the median COLIBRE MZRs \citep[][]{2014zndo.....11813N}. We only perform the fits in stellar mass bins where galaxies are sufficiently well resolved and the number of objects exceeds our fiducial threshold (see above). We show results for a representative COLIBRE box (L200m6) in \autoref{fig:mzr_fits}. While the adopted functional form captures the overall behaviour of the relation, the high-mass end at low redshifts is not well fit, reflecting the increasingly complex and non-monotonic shape of the MZR in this regime.

From the right panels of \autoref{fig:mzr_fits}, we see that although the best-fit $Z_0$ and $M_0$ vary non-linearly from $z=8$ to $z=0$, the general trend is that $Z_0$ increases over time as galaxies build up their metal content and the turnover shifts to higher metallicities. The low-mass slope $\gamma$ (filled circles connected by dashed curves) becomes less steep over time, implying a shallower power-law portion of the MZR at lower redshifts. The best-fit $\gamma \approx 0.35$ at $z=0$ is in very good agreement with that derived by \citet{2020MNRAS.491..944C} and \citet{2021ApJ...914...19S}. The parameter $\beta$ is comparatively less well constrained, as indicated by the larger uncertainties, but is consistent with that determined by both studies, as well as with \citet{2024ApJ...964...59L} who study massive galaxies at $z \approx 0.7$. Other observational and simulation-based works also find $\beta$ difficult to constrain and often adopt a fixed value \citep[e.g.,][]{2026ApJ..1000..109J}. The evolution of $\beta$, if any, is suggestive of a transition in the shape of the MZR: at high redshifts, the MZR is well described by a single power-law, which then transitions to a broken power-law as the turnover develops at later times. This transition appears to be in place by $z \approx 5$, towards the end of the epoch of reionization, but this could simply be due to the lack of very massive galaxies in the simulation volumes at these epochs. The MZR does not evolve beyond $z \geq 5$ in COLIBRE simulations, in agreement with previous works \citep[e.g.,][]{2024ApJ...967L..41M,2025ApJ...991L...4M}.

\begin{figure*}
\centering
\includegraphics[width=\textwidth]{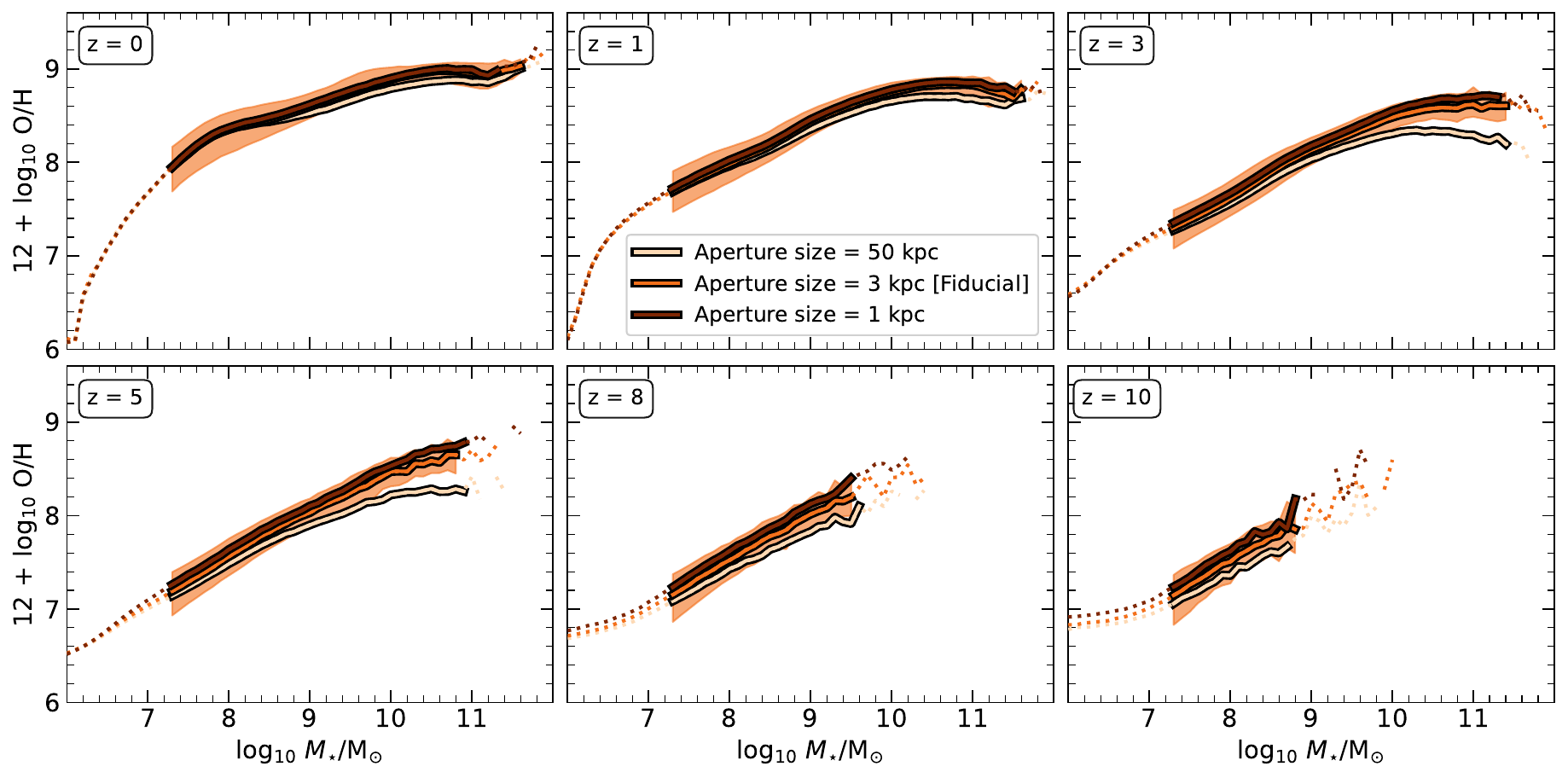}
\caption{Effects of using different aperture sizes to measure the gas-phase metallicity on the MZR in the COLIBRE L200m6 simulation. Median MZRs from these COLIBRE simulations are shown for three aperture sizes (50 kpc, 3 kpc, and 1 kpc), indicated by progressively darker shades of orange. The shaded band indicates the $16^{\rm{th}} - 84^{\rm{th}}$ percentile range in the fiducial case. The meaning of solid and dotted COLIBRE curves is the same as in \autoref{fig:mzr}.}
\label{fig:apertures}
\end{figure*}

We also fit the median MZRs with a single power-law \citep[SPL; e.g.,][]{2017MNRAS.465.1384C, 2018ApJ...859..175B,2025arXiv251200162R,2026AA...706A.165K}:
\begin{equation}
12 + \log_{10}\mathrm{O/H} = Z_0 + \gamma \log_{10}\left(\frac{M_{\star}}{10^{10}\rm{M_{\odot}}}\right)\,,
\label{eq:single-powerlaw}
\end{equation}
but we restrict the fit to $M_{\star} \leq 10^{10.4}\,\rm{M_{\odot}}$ at $z=0$ and $M_{\star} \leq 10^{10.2}\,\rm{M_{\odot}}$ at high-$z$, beyond which it is clear that a SPL model cannot capture the turnover. We find a qualitatively similar redshift evolution for the slope $\gamma$ (empty circles connected by dotted curves in the right panels of \autoref{fig:mzr_fits}). Both \citet{2021ApJ...914...19S} and \citet{2026ApJ..1000..109J} find virtually no evolution of $\gamma$ over $z \approx 0-3$ in the SPL fit (see also \citealt{2026MNRAS.tmp..397S}), whereas in COLIBRE $\gamma$ increases from 0.28 at $z=0$ to 0.4 at $z=3$ and then slowly saturates to 0.45. During the epoch of reionization, the best-fit $\gamma \approx 0.4$ we obtain is in good agreement with that estimated from observations \citep[e.g.,][]{2023NatAs...7.1517H,2024ApJ...976L..15C,2026MNRAS.546f2023R,2026arXiv260505327G}. The normalization $Z_0$, corresponding to the metallicity of a $10^{10}\,\rm{M_{\odot}}$ galaxy, does not evolve until $z \approx 5$, after which it increases monotonically $z=0$. It reflects the progressive enrichment of the ISM after several cycles of star formation and feedback, the declining importance of dilution by metal-poor gas accretion, and the ability of galaxies to retain more metals as they grow over time. Overall, these results indicate that the adopted functional form provides a useful, physically interpretable description of the evolving MZR, while also highlighting subtle discrepancies between observations and simulations.

\section{Discussion}
\label{s:discussion}
We have shown in \autoref{s:results} that COLIBRE is able to reproduce the MZR across cosmic time, including the turnover and saturation of the MZR, and the existence of bright, metal-rich galaxies at cosmic dawn. In this section, we explore how variations in modeling choices and key physical prescriptions impact the predicted MZR from COLIBRE, and discuss the broader implications of these dependencies for the evolution of metallicity in galaxies. The origin and evolution of the scatter in the MZR will be discussed in detail in a follow-up work (Burrafato et al. in preparation). To aid the comparison with model variations, we explicitly highlight the fiducial COLIBRE models we used in \autoref{s:results} in the legends of all plots in this section.

\subsection{Measurement choices}

\subsubsection{The effect of aperture size}
\label{s:apertures}

So far, we have used a constant aperture radius of physical size $3\,\rm{kpc}$ to calculate the oxygen abundances at each redshift from COLIBRE. It is important to assess the effects that aperture size can have on the MZR, since the aperture sizes used in observations vary significantly between different samples and across redshifts. For example, \citet[section 7]{2004ApJ...613..898T} showed that fixed aperture sizes lead to at most 0.11 dex differences in metallicity estimates in SDSS data. Motivated by these observational considerations, we use the L200m6 simulation to assess how aperture size affects the inferred MZR. Note that we continue to calculate stellar masses within $50\,\rm{kpc}$ apertures. Observational stellar mass estimates are typically derived from integrated photometry or SED fitting, which aim to capture the bulk of the galaxy light rather than being tied to a fixed physical aperture, and are therefore more comparable to large apertures. 

We test how much the median MZRs change if we use different aperture sizes to measure the oxygen abundance, finding no discernible differences for aperture sizes between $10-50\,\rm{kpc}$, so we do not discuss them further. However, there are some noticeable differences between apertures with sizes $\leq 3\,\rm{kpc}$ and larger ones, as we show in \autoref{fig:apertures}. We find that smaller apertures systematically yield higher median metallicities at fixed stellar mass, thereby raising the normalization of the MZR, with differences increasing with mass and reaching as high as $\approx 0.5\,\rm{dex}$ at high-$z$. This behaviour is expected, as smaller apertures preferentially sample the central regions of galaxies, which are typically more metal-enriched than the outskirts. The effect at the low-mass end is smaller since low-mass galaxies are also smaller in size. Comparing \autoref{fig:apertures} with \autoref{fig:mzr}, we see that COLIBRE MZRs with larger apertures better reproduce the observed population of the most massive galaxies at $z \approx 5$, whereas smaller apertures tend to overpredict the metallicity in massive galaxies at these epochs. We thus conclude that aperture size matters, particularly at high redshifts. The strongest effects, however, are limited to the most massive galaxies, which are more strongly affected by AGN feedback, as we discuss later in \autoref{s:agn}.

\begin{figure}
\centering
\includegraphics[width=\columnwidth]{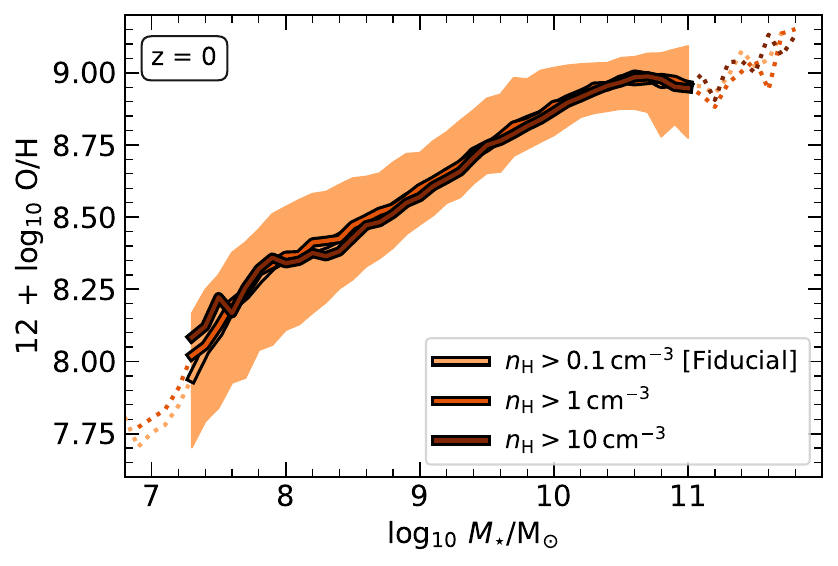}
\caption{The effect of using a different density cut to select gas particles to obtain oxygen abundances. Median MZRs at $z=0$ from COLIBRE simulations at m6 resolution and box size $100\,\rm{Mpc}$ (L100m6) are shown for the different density cuts, indicated by progressively darker shades of orange. The gas temperature threshold remains the same in all the cases ($T < 10^{4.5}\,\rm{K}$). The shaded band indicates the $16^{\rm{th}} - 84^{\rm{th}}$ percentile range in the fiducial case. The meaning of solid and dotted COLIBRE curves is the same as in \autoref{fig:mzr}. The results are similar across all redshifts.}
\label{fig:nhcut}
\end{figure}

\begin{figure}
\centering
\includegraphics[width=\columnwidth]{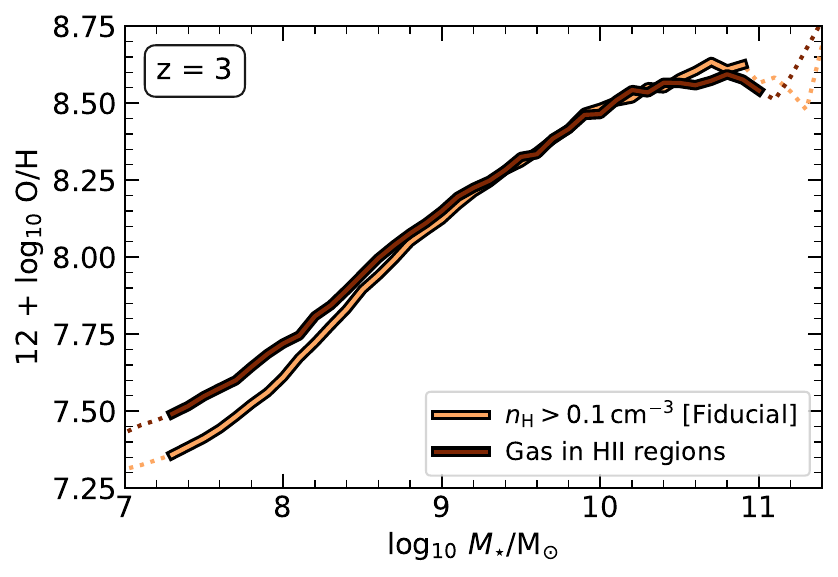}
\caption{The effect of using a different criteria to select gas particles to obtain oxygen abundances. The fiducial median MZR at $z=3$ is from the L100m6 COLIBRE simulation. The alternate MZR is created by using gas particles that are classified as being within H\textsc{II} regions irrespective of density. The gas temperature threshold remains the same in both cases ($T < 10^{4.5}\,\rm{K}$). The meaning of solid and dotted COLIBRE curves is the same as in \autoref{fig:mzr}. Although not shown here, quantitatively similar differences are found at $z > 3$.}
\label{fig:hiiregionendtime}
\end{figure}

\begin{figure*}
\centering
\includegraphics[width=\textwidth]{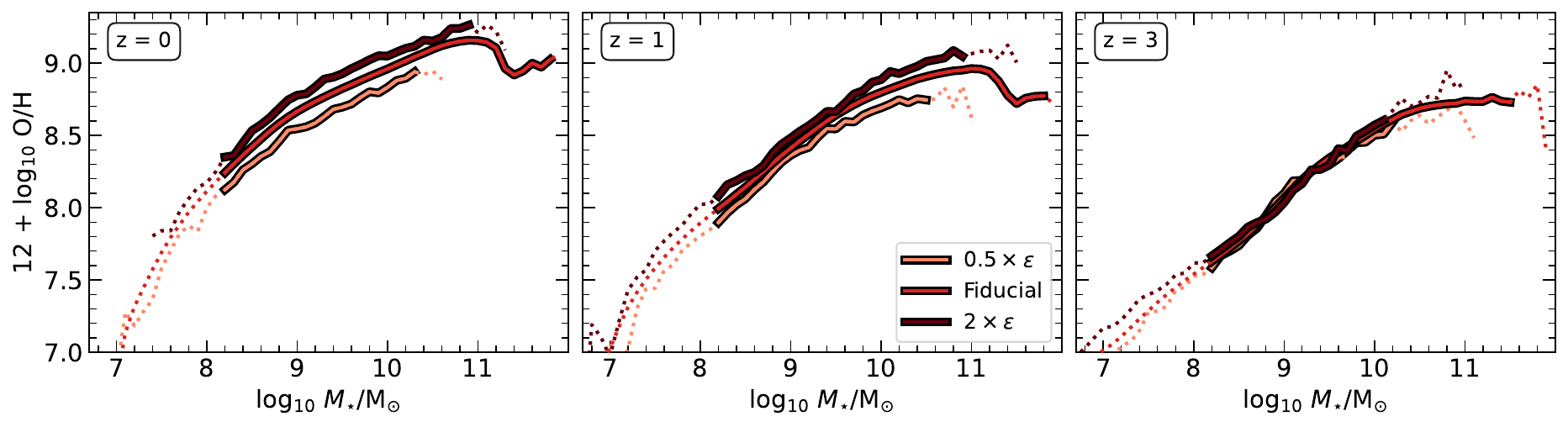}
\caption{Impact of variations in the fiducial star formation model implemented in COLIBRE on the median MZR, for a smaller (50 Mpc) box at m7 resolution (L050m7). The fiducial model is from the L400m7 simulation. $\epsilon$ is the star formation efficiency per freefall time, set to $0.01$ in the fiducial model \citep[][section 3.3]{2025arXiv250821126S}.}
\label{fig:sfe}
\end{figure*}

\begin{figure*}
\centering
\includegraphics[width=\textwidth]{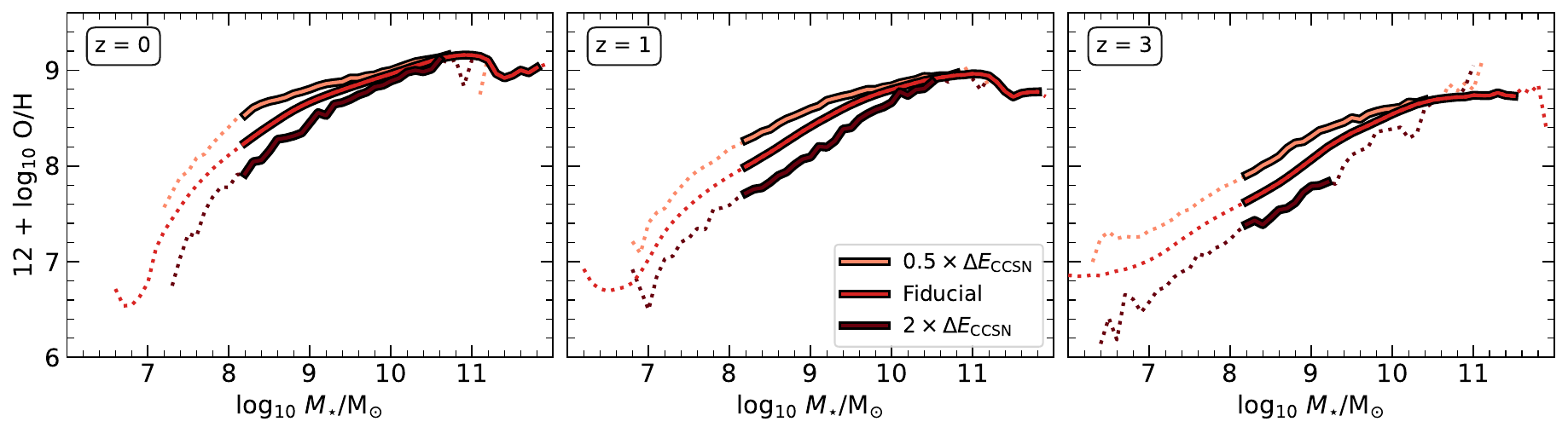}
\caption{Impact of variations in the fiducial supernova feedback model implemented in COLIBRE on the median MZR, for a smaller (50 Mpc) box at m7 resolution (L050m7). The fiducial model is from the L400m7 simulation. $\Delta E_{\rm{CCSN}}$ refers to the energy added to a star particle due to core collapse supernovae explosions within time $\Delta t$ \citep[][section 3.7]{2025arXiv250821126S}.}
\label{fig:ccsn}
\end{figure*}

\subsubsection{The effect of the ISM selection criterion}
\label{s:density}

In the analysis so far, we use gas particles that satisfy a fixed density and temperature threshold across cosmic time in all galaxies ($n_{\rm{H}} > 0.1\,\rm{cm^{-3}}$, $T < 10^{4.5}\,\rm{K}$), with the notion that such selection would preferentially trace star-forming gas and gas in H\textsc{ii} regions and is therefore appropriate for estimating the metallicity. However, high-resolution observations find that H\textsc{ii} regions often have $n_{\rm{H}} \gtrsim 1\,\rm{cm^{-3}}$ \citep[e.g.,][]{Osterbrock_1989,2001ApJ...556..121K,2009A&A...507.1327H,2016ApJ...816...23S,2024ApJ...974...34G}, so it is possible that our default selection includes diffuse, low-density ionized gas that can contaminate our metallicity measurements (an issue that also persists in observations that cannot resolve individual H\textsc{ii} regions -- \citealt{2017ApJ...850..136S,2019MNRAS.485..367K}). In fact, recent results from zoom-in simulations \citep[][]{2025arXiv251005201K} that explicitly follow ionized gas non-equilibrium chemistry \citep[][]{2022MNRAS.512..348K} suggest that ISM density plays a key role in modulating the slope of the MZR \citep[][]{2026OJAp....958199C}. 

To investigate this possibility, we modify our selection criteria by selecting gas particles with $n_{\rm{H}} > 1\,\rm{cm^{-3}}$ and $n_{\rm{H}} > 10\,\rm{cm^{-3}}$, respectively, while retaining the temperature threshold used above, to calculate the metallicity, and study the impact it has on the median COLIBRE MZRs. For this purpose, we use the L100m6 COLIBRE box and re-run the SOAP catalogues \citep[][]{2025JOSS...10.8252M} to generate the linear mass-weighted diffuse oxygen abundance for all gas particles based on the new criteria. As we show in \autoref{fig:nhcut}, adopting a higher density threshold to select gas particles while measuring the metallicity has little impact on the median MZR. We find this lack of dependence on the ISM density persists at high redshifts, and therefore omit those results for brevity.

Since COLIBRE follows hydrogen and helium species in non-equilibrium, one could in principle identify gas associated with H\textsc{ii} regions by selecting particles with a sufficiently high H$^+$ mass fraction, in combination with a limit on the gas temperature. However, even at m5 resolution, the H\textsc{ii} regions created by young stellar populations remain unresolved in COLIBRE, so such a selection could predominantly include diffuse ionized gas that is either not within the ISM of the galaxy, or does not originate from H\textsc{ii} regions. An alternative approach is to use gas particles explicitly tagged as belonging to H\textsc{ii} regions based on their distance to young stellar populations and the mass of the Strömgren sphere that such populations would create \citep[][section 3.6]{2025arXiv250821126S}. We apply this selection criterion in the L100m6 simulation, together with the same ceiling on the gas temperature as in the fiducial case ($T < 10^{4.5}\,\rm{K}$). We find that the resulting median MZRs are unchanged at $z < 3$. At higher redshifts ($z \geq 3$), this selection produces a modest metallicity enhancement of at most $0.1\,\rm{dex}$ in low-mass galaxies ($M_* < 10^{8.5}\,\rm{M_\odot}$), as shown in \autoref{fig:hiiregionendtime}. Therefore, we conclude that the median MZRs we present in this work are largely insensitive to the selection criteria we adopt to calculate the metallicity.

\subsection{Variations in the COLIBRE galaxy formation model}
\label{s:modelvariations}
In this section, we explore how the median MZRs evolve if we change key subgrid parameters that control star formation, and supernova and AGN feedback in the COLIBRE galaxy formation model. Since such variations are computationally expensive, they were only implemented in smaller simulation boxes that could be run down to $z=0$. Unless otherwise specified, we analyse simulations from the 50 Mpc box at m7 resolution (L050m7). A detailed examination of the impact of such variations on calibration benchmarks and other key galaxy scaling relations will be the subject of a future paper (Chaikin et al. in preparation); here, we focus on model parameters that could particularly impact the MZR.

\subsubsection{Impact of star formation efficiency}
\label{s:sfe}
The abundance of oxygen in the ISM is intricately linked to star formation which sets metal enrichment, and stellar feedback which regulates the amount of metals injected, dispersed and ejected from the ISM. The COLIBRE simulations use the volumetric \citet{1959ApJ...129..243S} law to convert gas that satisfies the star formation criterion into stars, which is modulated by the star formation efficiency per freefall time, $\epsilon$. The fiducial value of $\epsilon$ is set to 0.01, in line with the well established notion that on average one per cent of gas is converted into stars within a freefall time \citep[][]{2005ApJ...630..250K,2012ApJ...745...69K,2012ApJ...761..156F}. $\epsilon = 0.01$ also leads to excellent agreement between the predicted star formation efficiencies as a function of atomic and molecular gas surface densities from COLIBRE with observations \citep[][]{2025arXiv251211309L}. However, $\epsilon$ can reach higher values in denser ISM conditions, and can also be lower in H$_2$-poor environments \citep[e.g.,][]{2009ApJ...699..850K,2013MNRAS.433.1970F,2018MNRAS.477.2716K,2021ApJ...922L...3L,2025ApJ...987...12M}, so it is worth investigating how variations in $\epsilon$ impact star formation and subsequent metal enrichment. We therefore use two additional simulations of the L050m7 box, where $\epsilon$ was increased and decreased by a factor of 2 relative to the fiducial value.

\autoref{fig:sfe} shows the impact of varying $\epsilon$ on the median MZRs. As expected, we find that increasing (decreasing) $\epsilon$ leads to higher (lower) metallicities at fixed stellar mass by up to $0.3\,\rm{dex}$, since larger $\epsilon$ leads to more star formation per unit gas mass and freefall time, thereby leading to more metal enrichment. However, by comparing the normalized stellar mass distributions across the different runs, we find that these shifts are primarily driven by changes in the galaxy stellar mass rather than intrinsic changes in metallicity at fixed stellar mass. In particular, the $2\times\epsilon$ run produces a higher abundance of massive galaxies and fewer low-mass systems, with the opposite trend in the $0.5\times\epsilon$ run. Consequently, the slope of the MZR remains largely unchanged, and the apparent differences mainly reflect a redistribution of stellar masses across the population. By $z=3$, the differences in both the stellar mass distributions and the MZRs disappear, so we do not show median MZRs at higher redshifts. The apparent lack of sensitivity of the MZR to $\epsilon$ at high $z$ could imply that the gas (H\textsc{i} + H$_2$, since star formation can also occur in atomic gas in the COLIBRE model, especially at low metallicities -- see \citealt{2025arXiv251211309L}) depletion timescales become similar to the Hubble time at high $z$, but we do not find any clear distinction in the way depletion timescales vary as a function of stellar mass between low and high redshifts to support this scenario. It could also be that the gas accretion timescales become smaller than the Hubble time, in which case the gas reservoir is regulated primarily by the balance between cosmological inflows and stellar feedback-driven outflows, rather than by the local star formation efficiency.

\begin{figure*}
\centering
\includegraphics[width=\textwidth]{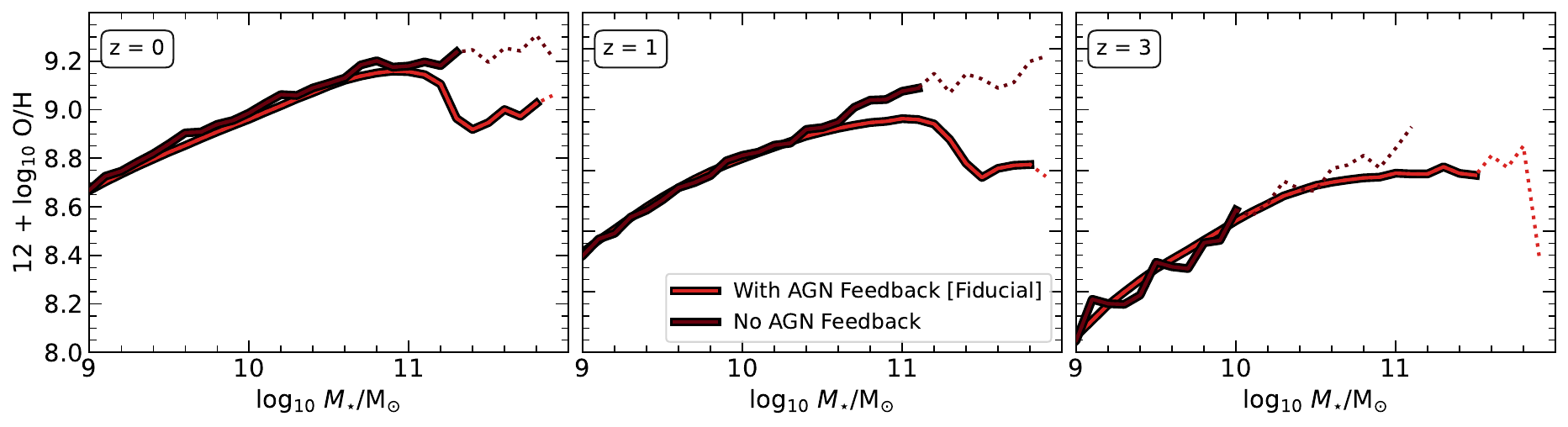}
\caption{Impact of AGN feedback on the median MZR in the COLIBRE simulations at m7 resolution. The fiducial model is plotted for the L400m7 box whereas the model without AGN feedback is for the L050m7 box. Notice that axes limits differ from those in other plots.}
\label{fig:noagn}
\end{figure*}

\begin{figure*}
\centering
\includegraphics[width=\textwidth]{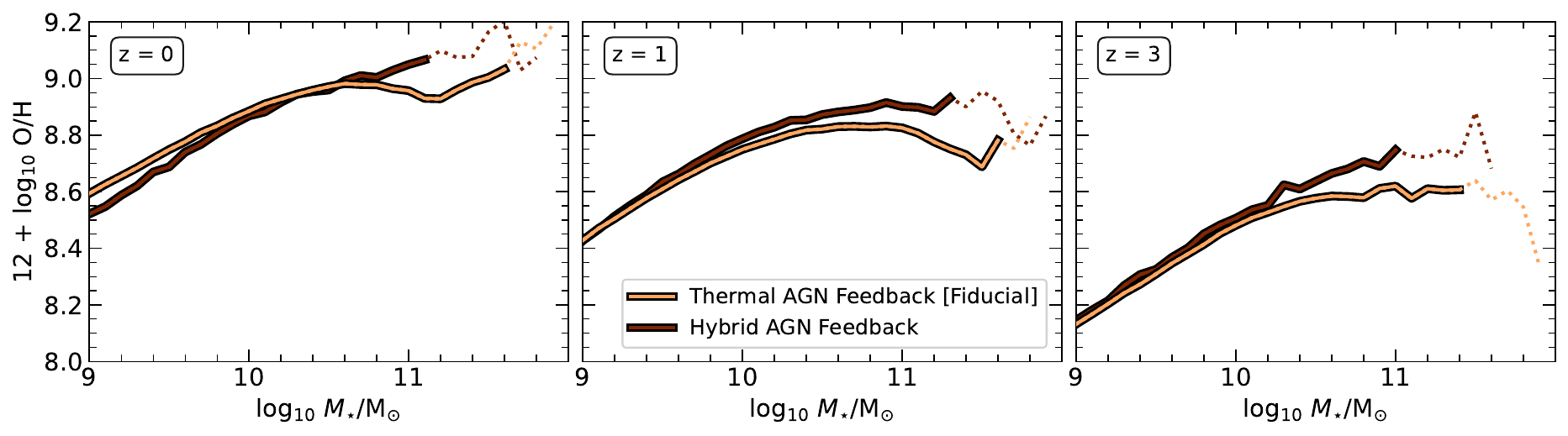}
\caption{Impact of the mode of AGN feedback (thermal versus hybrid) on the median MZR in the COLIBRE simulations of box size $100\,\rm{Mpc}$ at m6 resolution (L100m6). The fiducial thermal feedback model only injects thermal energy from AGN into the surroundings, whereas the hybrid feedback model injects both thermal and kinetic jet energy (see \autoref{s:agn} for more details). Notice that axes limits differ from those in other plots.}
\label{fig:agn}
\end{figure*}

\subsubsection{Impact of supernova feedback}
\label{s:snfeedback}

To model feedback from core-collapse supernovae (CCSNe), COLIBRE injects an energy of $10^{51}\,\rm{erg}$ per supernova event to neighboring gas particles, averaged over the stellar IMF and integrated over the simulation timestep $\Delta t$ \citep[][]{2012MNRAS.426..140D}. This energy is multiplied by a stellar birth pressure-dependent efficiency that is calibrated to the $z=0$ galaxy stellar mass function and galaxy sizes \citep[][]{chaikin25b}. The effective energy injection parameter is denoted by $\Delta E_{\rm CCSN}$ \citep[][equation 15]{2025arXiv250821126S}, whereas the number of gas particles that are heated due to supernova feedback is governed by how much energy is injected thermally (in the form of a temperature boost) and kinetically \citep[in the form of low-energy kicks,][]{2023MNRAS.523.3709C}. As for the star formation efficiency, variations in $\Delta E_{\rm{CCSN}}$ are expected as a function of local ISM properties, for example, due to supernova clustering, or supernovae exploding in environments with densities different than the mean ISM density \citep[e.g.,][]{2015ApJ...802...99K,2017MNRAS.465.2471G,2022ApJ...932...88O}. Note that by invoking a density-dependent temperature increment \citep[][equation 18]{2025arXiv250821126S}, COLIBRE partially accounts for variations in ISM density at the location of supernovae explosions.

\autoref{fig:ccsn} shows the impact of varying $\Delta E_{\rm{CCSN}}$ by a factor of two (at fixed stellar birth pressure) in the L050m7 COLIBRE box. The impact of supernova feedback on the MZR is more significant than the effect of $\epsilon$ -- stronger feedback tends to decrease the metallicity at the low-mass end as more metals are ejected out of the ISM, whereas weaker feedback tends to increase it (see also, \citealt{2017MNRAS.472.3354D}, on similar findings from the EAGLE simulations). The differences disappear for the massive ($M_* > 10^{10}\,\rm{M_{\odot}}$) galaxies at all redshifts, where AGN feedback has a larger influence as we show later in \autoref{s:agn}. While the $2\times\Delta E_{\rm{CCSN}}$ run produces fewer galaxies in the mass range $10^9 - 10^{10}\,\rm{M_{\odot}}$ and a correspondingly higher abundance of galaxies at lower masses, the resulting change in the MZR is not driven by this shift in the galaxy stellar mass. Instead, it arises because stronger supernova feedback systematically lowers the metallicity at fixed stellar mass relative to the fiducial run, and vice-versa. This is in contrast to the case of varying $\epsilon$, where the MZR response is largely mediated through changes in the stellar mass distribution. Consequently, the slope of the MZR is directly affected when supernova feedback is varied. This highlights the importance of small-scale baryonic processes in setting galaxy-scale chemical trends. In this context, high-resolution isolated galaxy patch (tall box) simulations that can resolve the launch and escape of metal-enriched multiphase outflows are particularly valuable \citep[e.g.,][]{2016MNRAS.456.3432G,2020ApJ...900...61K,2025MNRAS.539.1706V,2026MNRAS.549ag913V}. Such simulations can directly probe the origin of feedback-driven metal ejection and constrain how these processes coarse-grain into the effective $\Delta E_{\rm{CCSN}}$ parameterisation used in cosmological simulations like COLIBRE.

Comparing \autoref{fig:ccsn} with \autoref{fig:mzr}, we see that the shallower slope at the low-mass end at $z=1-3$ in the $0.5\times \Delta E_{\rm{CCSN}}$ run better reproduces MZRs from the JADES \citep[][]{2024AA...684A..75C}, MUDF \citep{2024ApJ...966..228R}, NGDEEP \citep[][]{2026arXiv260520810H} and A2744+SMACS surveys \citep[][]{2023ApJ...955L..18L}. Fits to the MZRs indicate that, in the $0.5\times \Delta E_{\rm{CCSN}}$ run, the SPL slope $\gamma$ (\autoref{eq:single-powerlaw}) evolves from 0.2 at $z=0$ to 0.4 at $z=3$. In contrast, the $2\times \Delta E_{\rm{CCSN}}$ run exhibits little evolution, with $\gamma \approx 0.50$ over the same redshift range. Taken together, these findings therefore suggest that stronger supernova feedback produces steeper MZRs whereas weaker supernova feedback produces shallower MZRs. This effect is in place at high redshifts as well, but we lack the necessary statistics due to the small volume to make meaningful comparisons at $z > 3$.

\subsubsection{Impact of AGN feedback}
\label{s:agn}
It is well established that AGN significantly regulate the growth of massive galaxies across cosmic time, and even indirectly impact low-mass galaxies via large-scale feedback, especially in galaxy clusters \citep[see reviews by][]{2014ARA&A..52..589H,2019ARA&A..57..467B}. Several previous works have also shown that AGN play a key role in setting the high-mass turnover of the MZR (e.g., via ejective feedback -- \citealt{2012MNRAS.422..215Y,2017MNRAS.472.3354D,2019MNRAS.486.2827D,2019MNRAS.484.2896T,2019MNRAS.484.5587T,2021MNRAS.504.4817V,2024MNRAS.52711043Y}; but see \citealt{2015MNRAS.448.1835T} for a contrasting view). To understand the impact of AGN feedback on the MZR, we compare the fiducial m7 simulation (L400m7) with a variant from the L050m7 box where AGN feedback is turned off. Since this is a small volume simulation, it does not contain a large number of massive galaxies. Nevertheless, \autoref{fig:noagn} shows a clear distinction at the massive end of the MZR between the two simulations: the MZR does not show a turnover in simulations without AGN feedback at $z \leq 1$, whereas the differences disappear at $z > 1$. Therefore, we conclude that AGN feedback is responsible for causing the downturn in the MZR at the massive end at low redshifts. At high redshifts, the impact of AGN on the MZR appears weaker; however, this may also reflect the fact that the turnover mass is not sufficiently sampled at high $z$.

The results we have presented so far are based on the `Thermal' model of AGN feedback in COLIBRE galaxies, where the feedback from the AGN is injected as thermal energy into particle(s) nearest to the central supermassive black hole \citep[following][]{2009MNRAS.398...53B}. The amount of energy injected is directly proportional to the accretion rate of the black hole, as well as the radiative efficiency. While this thermal-only feedback model is intended to capture the overall impact of AGN feedback in a phenomenological sense, without explicitly specifying the physical channel through which the energy couples to the surrounding gas, it does not directly model processes such as AGN jets, which can also transfer a significant amount of kinetic energy to the host galaxy and the circumgalactic medium (CGM, see reviews by \citealt{2017ARA&A..55..389T} and \citealt{2019ARA&A..57..467B}). To take this into account, a subset of COLIBRE simulations also include a `Hybrid' mode (thermal+kinetic) of AGN feedback, detailed in \citet{2026MNRAS.547ag324H}. Briefly, on top of the injection of thermal energy, the hybrid mode tracks the spin and accretion state of the accreting supermassive black hole to adjust the feedback based on the efficiency of wind and jet launching \citep[see also,][for similar approaches]{2019MNRAS.487..198G,2024MNRAS.532...60K}.

To test the impact of the AGN feedback models, we switch to using the L100m6 box for which there are simulations available with both thermal and hybrid AGN feedback. Unlike other model variants we have discussed above, the hybrid AGN feedback simulations were calibrated to match the AGN bolometric luminosity function at $z=0$ \citep[][]{2020MNRAS.495.3252S}, on top of the same set of observables used to calibrate the fiducial simulations \citep[][section 4.2]{2026MNRAS.547ag324H}. \autoref{fig:agn} shows the median MZRs at different redshifts of these simulations. This figure shows that the model with hybrid AGN feedback produces fewer massive ($M_* > 10^{11}\,\rm{M_{\odot}}$) star-forming galaxies across $z=0-3$. We find that while there is little difference at the low mass end (as expected), the median metallicity at the massive end is slightly elevated ($\lesssim 0.2\,\rm{dex}$) in the hybrid run as compared to the thermal run, but is within the error margin on metallicity estimates from observations. The turnover at the high-mass in COLIBRE is sensitive to the gas fraction (Burrafato et al. in  preparation) and hence to quenching, with more efficient AGN feedback creating a stronger turnover, as we saw above. From an observational point of view, we do not find any significant impact of the mode of AGN feedback on the MZR across cosmic time, and therefore do not show this comparison beyond $z > 3$.

In addition to the thermal versus hybrid model of AGN feedback, we also explored variations in two critical model parameters that govern AGN feedback, namely, the increase in temperature of the gas particles due to an AGN ($\Delta T_{\rm{AGN}}$, see equation 38 of \citealt{2025arXiv250821126S}), and the fraction of the intrinsic AGN luminosity that is coupled to the ambient gas ($\epsilon_{\rm{f}}$, see equation 34 of \citealt{2025arXiv250821126S}). The two parameters are however not independent of each other -- in the fiducial model $\Delta T_{\rm{AGN}}$ varies linearly with the black hole mass, which in turn is sensitive to $\epsilon_{\rm{f}}$. We find that such variations have a negligible impact on the median MZR across all redshifts, and therefore do not show them for brevity. This suggests that, within the explored parameter space, the detailed calibration of AGN feedback plays a subdominant role in shaping the MZR as compared to the presence and mode of AGN feedback.

\subsubsection{Impact of depletion of oxygen on dust}
\label{s:dustdepletion}
The depletion of oxygen onto dust grains can impact the MZR. It has been shown that due to depletion on dust, measurements in nearby H\textsc{ii} regions require a $\approx 0.1\,\rm{dex}$ correction to the oxygen abundance measurements \citep[][]{2010MNRAS.405.2651M,2010ApJ...724..791P}. Since COLIBRE includes a live dust model, we are in a unique position to directly trace oxygen depletion and study the impact of not taking dust depletion into account on the MZR. For this purpose, we define $f_{\rm{O,depl}}$, the fraction by which oxygen gets depleted onto dust grains (\textit{i.e.,} $1 + f_{\rm{O,depl}}$ specifies by what factor the MZR would shift if dust depletion is unaccounted for).

\begin{figure}
\centering
\includegraphics[width=\columnwidth]{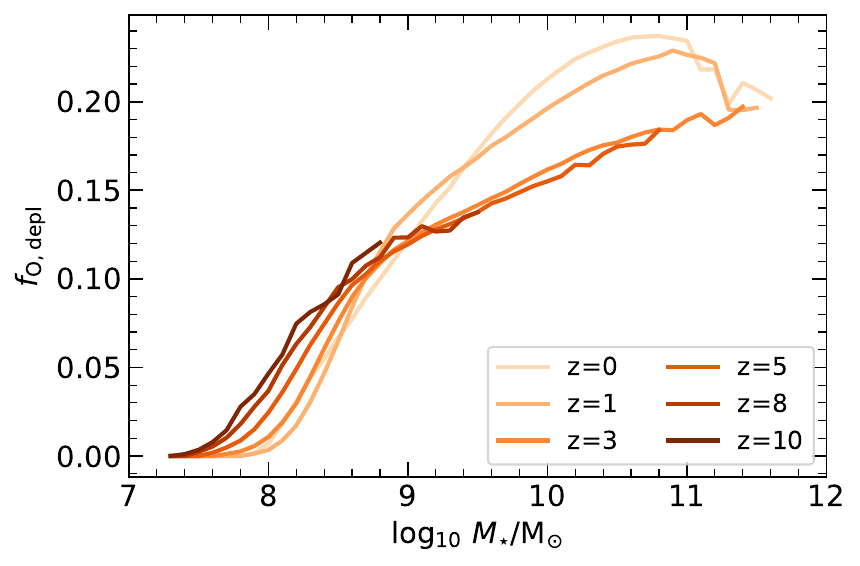}
\caption{Median fraction of oxygen that is depleted onto dust grains, $f_{\rm{O,depl}}$, as a function of galaxy stellar mass, at different redshifts in the COLIBRE L200m6 simulation.}
\label{fig:dustdepl}
\end{figure}

\autoref{fig:dustdepl} shows this fraction as a function of stellar mass at different redshifts for the fiducial L200m6 simulation. At all epochs, more massive galaxies exhibit higher depletion fractions, reflecting more efficient dust production and growth in their ISM. At the high-mass end, the depletion fraction increases steadily from high to low redshift, consistent with the gradual build-up of dust reservoirs as galaxies evolve.

\begin{figure}
\centering
\includegraphics[width=\columnwidth]{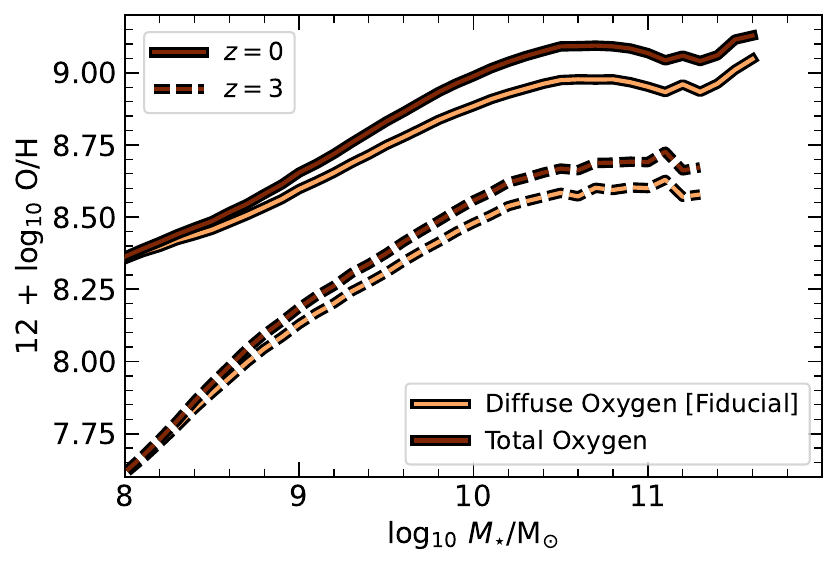}
\caption{MZR for the L200m6 COLIBRE simulation at $z=0$ (solid) and $z=3$ (dashed), showing the impact of ignoring oxygen depletion onto dust grains. The darker shade of orange corresponds to the total oxygen abundance (gas+dust) whereas the lighter shade corresponds to the fiducial MZR that only takes the abundance of oxygen in the gas-phase into account.}
\label{fig:mzr_dustdepl}
\end{figure}

At lower stellar masses, however, the redshift evolution differs. Below $\log_{10}(M_{\star}/\mathrm{M}_{\odot}) \approx 8.5$, the depletion fraction decreases from high to low redshift, implying that, at fixed stellar mass, galaxies at earlier times contain a larger fraction of their oxygen locked in dust. In COLIBRE galaxies, this transition mass corresponds to the regime where the dust-to-metal ratio (DTM) shifts from a rising trend with stellar mass to a saturated plateau, set by the balance between grain growth and destruction (Vijayan et al. in preparation). In the low-mass regime, the DTM is sensitive to the density of the ISM, with higher densities promoting more efficient grain growth. Since galaxies at higher redshift tend to have denser ISM conditions \citep[e.g.,][]{2026OJAp....958199C}, they achieve higher DTMs at fixed stellar mass, reaching the saturation regime at lower masses. While this naturally explains the apparent reversal in the redshift dependence of the depletion fraction at the low-mass end, it could be sensitive to resolution if the dense ISM is not sampled by a sufficient number of particles.

\autoref{fig:mzr_dustdepl} shows the resulting MZRs for the L200m6 simulation when dust depletion is not taken into account. We present results at two representative redshifts, noting that the behaviour is similar at other epochs. Neglecting depletion of oxygen onto dust grains raises the median MZR by $\approx 0.1\,\rm{dex}$ at the high-mass end, which however lies within the systematic uncertainties of the observations. Despite variations in $f_{\rm{O,depl}}$ above, we thus do not find a large impact of dust depletion on the MZR. However, it is expected to leave a more significant imprint on elements other than oxygen, as well as the DTM ratio (Vijayan et al. in preparation), and should therefore be included when making detailed comparisons to observational data.

\section{Summary and Outlook}
\label{s:summary}
In this work, we present the gas-phase mass-metallicity relation (MZR) in galaxies across cosmic time as predicted by the COLIBRE cosmological simulations \citep[][]{2025arXiv250821126S,chaikin25a}. By explicitly modeling the cold ISM (down to $10\,\rm{K}$) and dust grains, while achieving comparable or higher resolution and similar or larger simulation volumes than previous generations of cosmological simulations, COLIBRE represents a significant step forward in modeling ISM properties in a cosmological context. We use the three COLIBRE fiducial simulations with particle masses, for both baryons and dark matter, of $\sim 10^5\,\rm{M_{\odot}}$ (m5), $\sim 10^6\,\rm{M_{\odot}}$ (m6), and $\sim 10^7\,\rm{M_{\odot}}$ (m7), and the largest available simulation box at each resolution and redshift (see \autoref{tab:tab1}) to produce the median MZR of star-forming galaxies from $0 \leq z \leq 15$. COLIBRE MZRs show excellent convergence with both the simulation volume and resolution. The predicted MZRs are robust against the choice of apertures used to measure the metallicity, except for the most massive galaxies at $z \approx 3-5$. We compare COLIBRE MZRs with a vast compilation of observational measurements and predictions from other simulations. Our key findings are as follows:

\begin{enumerate}
    \item The COLIBRE simulations reproduce the trends seen in the MZR across cosmic time, showing that the MZR is already in place at cosmic dawn ($z = 10$, \autoref{fig:mzr}). Unlike other existing simulations, COLIBRE contains a statistical sample of resolved galaxies across the entire stellar mass range probed observationally at all redshifts (\autoref{fig:mzr_sims}), and can therefore be used to derive robust conclusions on the predicted evolution of the MZR.
    \item The median trend in gas-phase metallicity as a function of redshift in COLIBRE galaxies is in good agreement with the data at most stellar masses across all redshifts (\autoref{fig:redshift_evolution_all}), but COLIBRE simulations overpredict the metallicity of galaxies with $M_{\star} \approx 10^{10}\,\rm{M_{\odot}}$ at $z \approx 5-8$.
    \item The slope of the COLIBRE MZRs, whether fitted with broken or single power-law models, suggests that the MZR did not significantly evolve from $z > 10$ down to the end of the epoch of reionization ($z \approx 5$). After $z \approx 3$, the slope became shallower with time at low redshifts (\autoref{fig:mzr_fits}). 
    \item Variations in the star formation efficiency per free-fall time lead to modest differences in the MZR at $z \leq 1$, with higher efficiencies yielding higher metallicities (\autoref{fig:sfe}).
    \item Stronger supernova feedback leads to a steeper slope of the MZR at the low-mass end, and vice-versa, highlighting the central role of supernova feedback in regulating galaxy metallicity (\autoref{fig:ccsn}).
    \item In the absence of AGN feedback, the MZR saturates at high stellar masses but does not exhibit a downturn (\autoref{fig:agn}). Including a hybrid (thermal + kinetic jet) mode of AGN feedback slightly raises the upper end of the MZR. 
    \item The effects of dust depletion on the MZR are modest (up to $\approx 0.2$~dex; \autoref{fig:mzr_dustdepl}). COLIBRE reveals a redshift reversal in dust depletion for low-mass galaxies driven by the fact the Universe is denser at early epochs (\autoref{fig:dustdepl}).
    
\end{enumerate}

However, we note that the observational data suffers from significant uncertainties due to inhomogeneous samples that use a variety of metallicity calibrations to construct the MZR. As such, further efforts are needed to measure the metallicities in a consistent way using the direct method for a large sample of galaxies to enable a more robust comparison with theory \citep[e.g.,][]{2020MNRAS.491..944C,2026ApJ..1000..109J,2026arXiv260102463S}. 

It would be useful to post-process COLIBRE galaxies with photoionization modeling codes such as \texttt{CLOUDY} \citep[][]{Ferland_1998} and radiative transfer codes like \texttt{SKIRT} \citep[][]{2015A&C.....9...20C} or \texttt{SUNRISE} \citep[][]{2006MNRAS.372....2J} to generate mock emission-line data to enable a more apples-to-apples comparison with gas-phase metallicity observations \citep[e.g.,][]{2015MNRAS.454.2381G,2023MNRAS.526.3504H,2023MNRAS.526.3610H,2023ApJ...953..140N,2024ApJ...972..113G,2026OJAp....958199C}, and investigate whether the intrinsic MZRs presented in this work differ systematically from those inferred using observational diagnostics. In a follow-up work, we will present a study of the scatter in the COLIBRE MZRs across cosmic time, and residual correlations with SFR, gas fractions, outflow rates, etc. (Burrafato et al. in preparation). In addition to oxygen abundances, it will also be interesting to explore variations in C/O and N/O in star-forming galaxies over cosmic time, which provide important insights on the impact of stellar nucleosynthesis on galactic chemical evolution \citep[e.g.,][]{2017MNRAS.466.4403N,2019ApJ...874...93B,2020ApJ...900..179K}. In addition to the integrated galaxy metallicities, the COLIBRE simulations can also be used to explore abundance gradients to provide a resolved view of the baryon cycle in galaxies.

Beyond the inhomogeneity of the data and the need for consistent abundance determinations, our findings raise a broader question: what (additional) observations could be most critical for constraining theoretical MZRs and discriminating between the various physical mechanisms proposed in the literature? We present four key avenues to help answer this question:

\begin{enumerate}
    \item As we show in \autoref{fig:mzr_sims}, different simulations diverge significantly at the low-mass end of the MZR across all redshifts, highlighting the need for robust metallicity measurements in galaxies with $M_{\star} \lesssim 10^8\,\rm{M_{\odot}}$. In particular, the data we compile from the literature indicates that even at low redshifts ($z \leq 3$), there is a lack of statistically meaningful samples in this mass regime with reliable metallicity estimates. 
    
    \item At the opposite end, we observe substantial scatter in the metallicities of massive galaxies ($M_{\star} \simeq 10^{10},\rm{M_{\odot}}$) at $z > 4$, with current data suggesting that most simulations tend to overpredict their metallicities. At $M_* \simeq 10^{11}\,\rm{M_{\odot}}$, there are only two measurements available at $z > 3$. Given that AGN feedback is predicted to play a central role in regulating metal enrichment in these systems \citep[see also,][figure 13]{2017MNRAS.472.3354D}, improved constraints on the metallicities of massive galaxies at high redshift would provide valuable insight into the impact of AGN-driven processes on the baryon cycle.
    
    \item Thanks to live dust modeling, COLIBRE has ushered in an era wherein the dust-to-gas and dust-to-metal ratios can be measured in situ in cosmological simulations \citep[see also,][]{2026arXiv260221790R}, underscoring the need for equally reliable observational constraints on these quantities to enable meaningful comparisons \citep[e.g.,][]{2026MNRAS.545f1897A}. 
    
    \item Finally, as we illustrate in \autoref{fig:mzr_contours}, while current observations have begun to probe the most massive galaxies at $z > 10$, the underlying galaxy population at these epochs is likely dominated by lower-mass systems. Expanding spectroscopic samples toward these fainter, lower-mass galaxies -- through deeper surveys and gravitational lensing -- will be crucial for establishing the earliest phases of the MZR in the first few 100 Myr after the Big Bang.

\end{enumerate}

\section*{Acknowledgements}
We thank Lucie Rowland for providing feedback on an earlier version of this paper, Ryan Sanders for insightful discussions on the MZR, and Gauri Kotiwale for sharing their stacked measurements. We are grateful to Alex Garcia for making the data used to construct the MZRs from other cosmological simulations publicly available, and for supplying additional data upon request. PS is supported by the Leiden University Oort Fellowship and the International Astronomical Union -- Gruber Foundation (TGF) Fellowship. ABL acknowledges support by the Italian Ministry for Universities (MUR) program `Dipartimenti di Eccellenza 2023-2027' within the Centro Bicocca di Cosmologia Quantitativa (BiCoQ), and support by UNIMIB’s Fondo Di Ateneo Quota Competitiva (project 2024-ATEQC-0050). EC acknowledges support from Science and Technology Facilities Council (STFC) consolidated grant ST/X001075/1. CSF acknowledges support from European Research Council (ERC) Advanced Grant DMIDAS (GA 786910). JAH acknowledges support from the ERC Consolidator Grant 101088676 (``VOYAJ''). FH acknowledges funding from the Netherlands Organization for Scientific Research (NWO) through research programme Athena 184.034.002. SP acknowledges support by the Austrian Science Fund (FWF) through grant-DOI: 10.55776/V982. This work used the DiRAC@Durham facility managed by the Institute for Computational Cosmology on behalf of the STFC DiRAC HPC Facility (\url{www.dirac.ac.uk}). The equipment was funded by BEIS capital funding via STFC capital grants ST/K00042X/1, ST/P002293/1, ST/R002371/1 and ST/S002502/1, Durham University and STFC operations grant ST/R000832/1. DiRAC is part of the National e-Infrastructure. We acknowledge using the following softwares: \texttt{SWIFT} \citep{2024MNRAS.530.2378S}, \texttt{SwiftSimIO} \citep{2020JOSS....5.2430B}, \texttt{SwiftGalaxy} \citep{2025JOSS...10.9278O}, \texttt{PARTRIDGE} (Huško et al. in preparation), \texttt{SOAP} \citep{2025JOSS...10.8252M}, and \texttt{cmasher} \citep{2020JOSS....5.2004V}.

\section*{Data Availability}
The data and scripts underlying all the plots presented in this work are available at \url{https://github.com/psharda/colibre_mzr_sharda26}. The COLIBRE simulation data will eventually be made publicly available, although we note that the data volume (several petabytes) may prevent us from simply placing the raw data on a server. In the meantime, people interested in using the simulations are encouraged to contact the COLIBRE PI, and check the COLIBRE website for updates (\url{https://colibre-simulations.org}). A public version of the \texttt{SWIFT} code \citep{2024MNRAS.530.2378S} is available at \url{http://www.swiftsim.com}. The COLIBRE modules implemented in \texttt{SWIFT} will be made publicly available after the public release of the simulation data.



\bibliographystyle{mnras}
\bibliography{references} 


\appendix

\section{Convergence of the MZR with the box size}
\label{s:app_boxsize}
In this Appendix, we assess the convergence of the MZR with the box size of the COLIBRE simulations across cosmic time. To this end, we compare simulations run at fixed m6 resolution but with different box sizes (200 cMpc, 100 cMpc, and 25 cMpc, respectively), allowing us to test whether the median MZR is sensitive to simulation volume.

\begin{figure*}
\centering
\includegraphics[width=\textwidth]{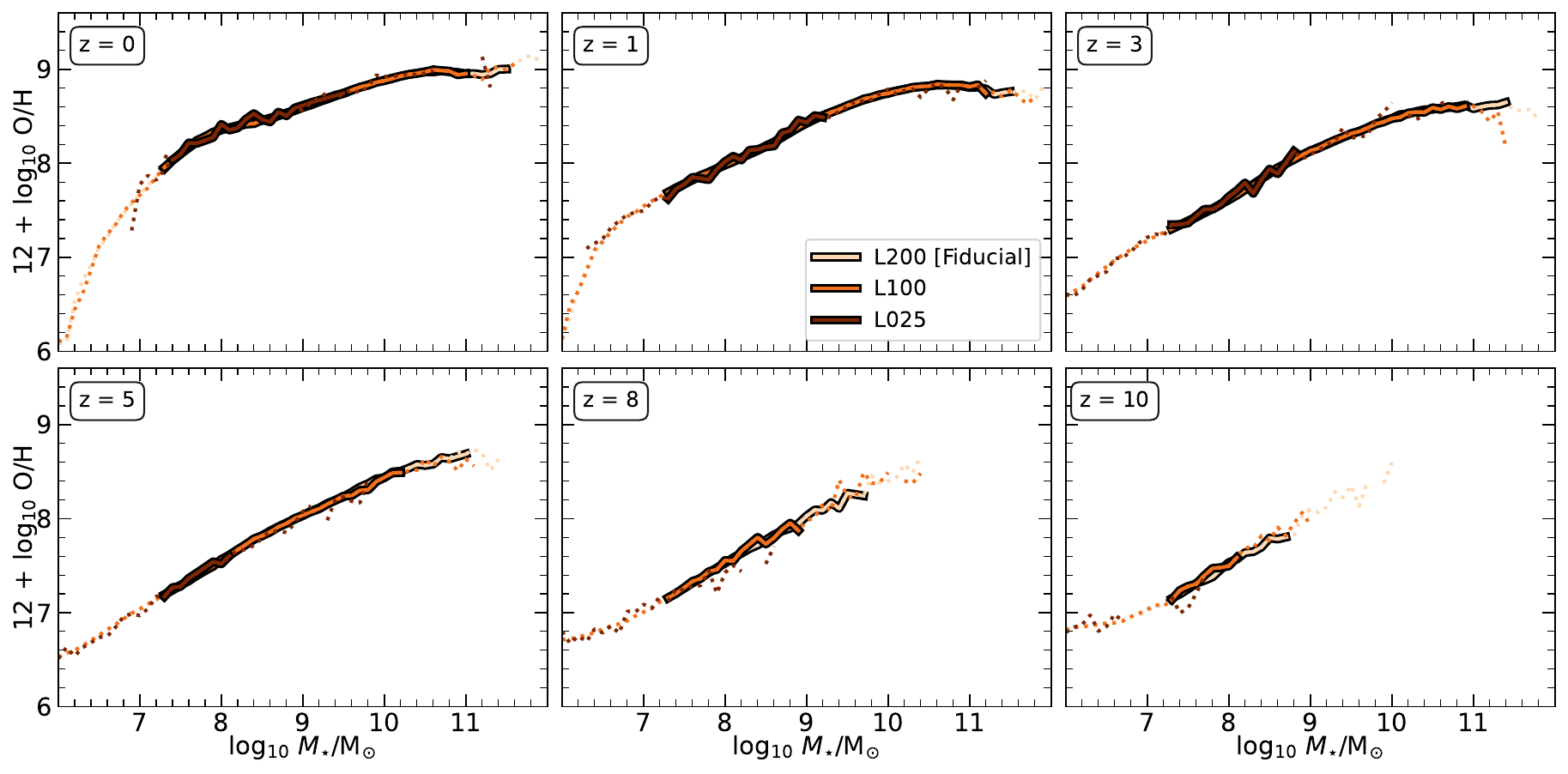}
\caption{Convergence tests for the median MZR at fixed (m6) resolution and varying box sizes in the COLIBRE simulations. Median MZRs for three box sizes (200 Mpc, 100 Mpc and 25 Mpc) are shown. The meaning of the solid and dotted curves is the same as that in \autoref{fig:mzr}.}
\label{fig:convergence}
\end{figure*}

\autoref{fig:convergence} shows that, at fixed resolution, the MZR is fully converged across all redshifts over the stellar mass range common to the different box sizes. While such agreement is expected at $z=0$, where galaxy samples are large and the majority of stellar mass bins are well populated, it is particularly encouraging that this convergence persists to the highest redshifts we consider here. This demonstrates that the inferred MZR is robust to volume effects over the stellar mass range sampled in common by the different simulations, despite the increasingly limited number of galaxies at early times. However, different simulation volumes probe different stellar mass ranges, particularly at high redshift where the most massive galaxies become increasingly rare, and therefore features such as the high-mass turnover may not always be sampled in smaller volumes.

\section{The effect of random errors in stellar mass on the MZR}
\label{s:app_mstar_random}
In the main text, we mimic the effect of uncertainties in stellar mass measurements by introducing a redshift-dependent lognormal scatter in $M_{\star}$ that increases at high redshifts (up to a ceiling of 0.3 dex; see \autoref{eq:scatter}). However, the magnitude of this scatter is not well constrained, so in this Appendix, we study the impact of varying the magnitude of the scatter by systematically varying the standard deviation $\sigma_{M_*}$ from $0.0$ to $0.5\,\rm{dex}$ \citep[see also,][]{2025arXiv250821126S,chaikin25b,2026arXiv260326200L}. We only show the results for the L200m6 simulation, noting that the results from other runs are similar.

\begin{figure*}
\centering
\includegraphics[width=\textwidth]{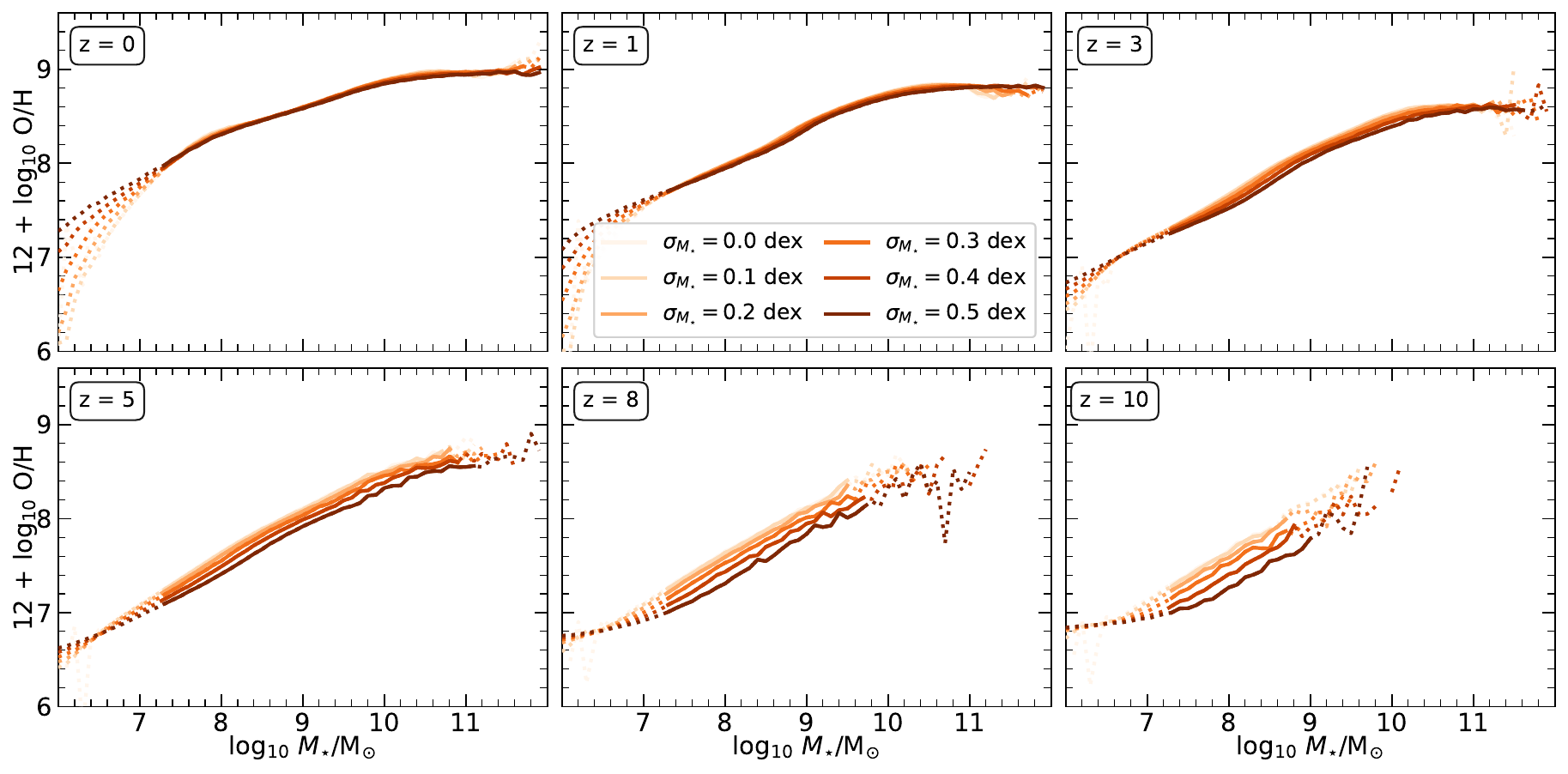}
\caption{Impact of random scatter in stellar masses on the median MZR in the COLIBRE simulations. Increasingly darker shades of orange correspond to lognormal scatter of up to 0.5 dex in the stellar mass. Shaded region denotes the $16^{\rm{th}} - 84^{\rm{th}}$ percentile range for the MZR with $\sigma_{M_*} = 0$. Only the results for the m6 resolution are shown. The meaning of solid and dotted COLIBRE curves is the same as in \autoref{fig:mzr}.}
\label{fig:randomscatter}
\end{figure*}

\autoref{fig:randomscatter} shows the resulting median MZRs when we vary $\sigma_{M_*}$. The shaded region denotes the $16^{\rm{th}} - 84^{\rm{th}}$ percentile range for the MZR with no scatter. While the differences are negligible at low redshifts, at $z > 3$, the increase in scatter corresponds to a decrease in the median metallicity at fixed stellar mass. If we did not take this scatter into account, the COLIBRE MZRs at high $z$ would predict somewhat elevated metallicities at fixed stellar mass. Overall, we find that the effects of uncertainties in the stellar mass can be different at different redshifts and $M_*$, and approach the magnitude of systematic uncertainty in the metallicity measurements.

\bsp    
\label{lastpage}
\end{document}